\documentclass[12pt]{article}

\usepackage{amsthm,amssymb,amsmath}
\usepackage{blindtext}
\usepackage{setspace}
\usepackage[margin = 1in]{geometry}
\usepackage{comment}
\usepackage[citecolor=blue,colorlinks]{hyperref}
\usepackage{booktabs}
\usepackage{graphicx}
\usepackage{multirow}
\usepackage{float}
\usepackage{fancyhdr}

\usepackage{natbib}

\title{Bayesian Causal Machine Learning for Cure Models}
\author{
  Antonio R. Linero\thanks{\texttt{antonio.linero@austin.utexas.edu}},
  F. Javier Rubio\thanks{\texttt{f.j.rubio@ucl.ac.uk}},
  and
  Piyali Basak\thanks{\texttt{piyali.basak@merck.com}}
}

\newcommand{\BartCure}{\texttt{BartCure}}
\newcommand{\Bernoulli}{\operatorname{Bernoulli}}

\newcommand{\CS}{\text{CS}}
\newcommand{\DR}{\text{DR}}
\newcommand{\E}{\mathbb E}
\newcommand{\expit}{\operatorname{expit}}
\newcommand{\Gam}{\operatorname{Gam}}
\newcommand{\iid}{\stackrel{\textnormal{iid}}{\sim}}
\newcommand{\Leaves}{\operatorname{Leaves}}
\newcommand{\logit}{\operatorname{logit}}
\newcommand{\NDS}{\text{NDS}}
\newcommand{\Normal}{\operatorname{Normal}}

\newcommand{\Poisson}{\operatorname{Poisson}}

\newcommand{\Reals}{\mathbb R}

\newcommand{\sM}{\mathcal M}
\newcommand{\Tree}{\mathcal T}
\newcommand{\Uniform}{\operatorname{Uniform}}
\newcommand{\Var}{\operatorname{Var}}

\newcommand{\Ft}{\widetilde{F}}
\newcommand{\ft}{\widetilde{f}}
\newcommand{\St}{\widetilde{S}}
\newcommand{\htt}{\widetilde{h}}

\newtheorem{proposition}{Proposition}

\begin{document}

\maketitle


\begin{abstract}
  In survival studies, treatments can benefit patients through different mechanisms: a treatment may increase the probability of being \emph{cured} or delay failure among patients who are not cured. Quantifying which mechanism is dominant, and whether it varies across subpopulations, is clinically important, yet there is limited work in the causal machine learning literature addressing this problem. Standard causal survival learners target finite-horizon survival or restricted mean survival time, while many cure models capture cure structures without estimating causal effects. In this work, we define meaningful causal effects in the presence of a cured subpopulation and introduce \BartCure, a Bayesian causal machine learning approach for estimating them. The causal effects we recommend decompose the causal effect on restricted mean survival time into a \emph{stochastic cure} and \emph{stochastic latency} component, and we relate these new effects to both stochastic intervention effects and causal effects in principal strata. In simulations, \BartCure\ is competitive for estimating average effects and is especially effective at conservatively detecting the direction of treatment-effect heterogeneity. We apply \BartCure\ to estimate average and subgroup causal effects and to identify treatment effect heterogeneity in the CALGB 40101 breast cancer trial.

\end{abstract}

\doublespacing

\section{Introduction}

In many clinical time-to-event studies, a non-negligible subset of patients never experience the event of interest because they are no longer susceptible to failure. For example, in breast cancer studies, patients may be successfully treated and enter long-term remission, effectively being ``cured'' of their cancer. In these settings, standard survival models can blur two distinct features of the outcome process: whether a patient remains at risk at all, and how quickly events occur among those who are still susceptible. One approach for addressing this setting is the use of a \emph{cure rate} model, which allows the event time to have a positive probability of being infinite (see \citealp{othus2012cure} and \citealp{peng2014cure} for detailed treatments). This additional modeling is useful because it isolates the ways in which a treatment may provide benefit; for example, a treatment might shorten survival among individuals who are not cured while simultaneously increasing the probability of being cured.


The interpretation of cure models as distinguishing different ways in which a treatment may affect patient outcomes naturally suggests a causal perspective, yet there has been surprisingly little work in this direction. Comparing survival times only among observed failures across treatment groups is insufficient, because the population of individuals who experience an event under treatment may differ from the population who experience an event under control. Recently, \citet{wang2024causal} used principal stratification \citep{frangakis2002principal} to define causal effects within the subpopulation of patients who are ``always uncured'', \textit{i.e.}, patients who would experience an event under either treatment arm. However, identification of these effects requires additional monotonicity assumptions on the treatment effect, and in many clinical trials, including the noninferiority CALGB 40101 \citep{shulman2014comparison} trial analyzed in this paper, such assumptions may be difficult to justify.

In this work, we define causal estimands tailored to the cure rate setting and develop flexible methods for estimating them. Our main contributions are summarized as follows:
\begin{enumerate}
\item We define useful parameters in the cure-rate setting to target using causal machine learning, including effects that decompose restricted mean survival time effects into effects attributable and not attributable to cure process.
\item We introduce \BartCure\ as an effective method for obtaining estimates of population and conditional-average variants of these estimands. We then evaluate \BartCure\ on both synthetic and real data, finding that it compares very favorably with other causal machine learning methods for estimating effects and detecting treatment effect heterogeneity.
\item We provide tools to use with \BartCure\ for interpreting the fitted model to understand the different sources of treatment effect heterogeneity.
\end{enumerate}
Natural causal estimands in settings with a cured population are (i) the effect on the probability of an individual being cured, and (ii) the causal effect on the restricted mean survival time (RMST, \citealp{royston2013restricted}). To isolate the effect of the treatment on outcomes \emph{removing} the effect of the treatment on cure, we also define a \emph{latency effect} that decomposes the RMST effect into a part depending on the cure effect and a part depending on the latency effect; we also define a \emph{stochastic latency effect} that we argue is more interpretable. We show that the latency effect is a well-defined contrast of distributions of potential outcomes on the same population, and therefore defines a causal effect; an advantage of the latency effect over the proposals of \citet{wang2024causal} is that it can be estimated with minimal assumptions beyond those typically made in causal survival analysis.


As a particular estimation strategy, we introduce \BartCure, a promotion-time cure rate model \citep{chen1999new,yakovlev1996stochastic} based on Bayesian additive regression trees (BART, \citealp{chipman2010bart}), and provide evidence that it performs favorably relative to other causal machine learning techniques. \BartCure\ is fully nonparametric in the sense that it can capture arbitrary covariate-dependent variation in the survival function over time. We provide a rigorous evaluation of its estimation accuracy and interval coverage for average and conditional average causal effects, as well as its ability to detect meaningful treatment effect heterogeneity. We believe that both the model itself and the novel data augmentation algorithm used to fit it are of independent methodological interest. An advantage of \BartCure, and Bayesian machine learning more broadly, is the availability of direct uncertainty quantification through the posterior distribution. Empirically, these methods have demonstrated strong performance: BART has been a highly competitive prediction method since its introduction, and BART-based causal estimators have performed well in the ACIC data analysis competition \citep{dorie2019automated,thal2023causal}. More recently, \citet{kabata2025quantifying} showed that the strong empirical performance of BART-based methods extends to causal survival analysis.

We apply our methodology to a large clinical trial (CALGB 40101) that aimed to establish the noninferiority of the treatment paclitaxel (T, $A_i = 1$) for breast cancer relative to a baseline treatment cyclophosphamide+doxorubicin (CA, $A_i = 0$); because T has milder side effects than CA, this would allow T to be used as a new standard of care. To facilitate interpretation of causal effect estimates, we also show how to summarize the \BartCure\ posterior.
Our analysis shows that there is limited evidence of treatment effect heterogeneity in the relative performance of these treatments. We find weak evidence that the gap in survival is higher in older patients.
We also see mild differences between CA and T over shorter time horizons, with the main treatment effect being that CA cures more individuals as opposed to lengthening the survival time of uncured individuals.

\subsection{Related Work}

This work contributes to a broader literature on causal machine learning in survival analysis. Beyond additional flexibility, we believe an important advantage of \BartCure\ is that it reduces the ``researcher degrees of freedom'' \citep{simmons2011falsepositive} associated with analysts manually specifying nonlinearities and interaction effects. Several recent methods target treatment effect heterogeneity in time-to-event settings without explicitly modeling cure, including targeted learning approaches for conditional survival effects \citep{zhu2020targeted}, Bayesian accelerated failure time models for individualized treatment effects \citep{henderson2020individualized}, causal survival forests \citep{cui2023estimating}, and deep representation learning methods for treatment-specific hazards \citep{curth2021survite}. These methods provide flexible tools for handling censoring and treatment effect heterogeneity, but they generally treat survival as finite rather than explicitly modeling a cured subpopulation. 

In the context of BART, several survival models for causal inference have been proposed. The best-known approach, introduced by \citet{henderson2020individualized}, models the logarithm of the survival time as $\log T_i(a) = r(a, X_i) + W_i$, where the error distribution $W_i$ is modeled using a Dirichlet process mixture model \citep{escobar1995bayesian}. This additive specification for $T_i(a)$ induces an \emph{accelerated failure time} (AFT) structure. The method is implemented in the \texttt{AFTrees} package in \texttt{R}. In extensive simulation studies, \citet{kabata2025quantifying} showed that \texttt{AFTrees} compares very favorably with causal survival forests in both estimating heterogeneous causal effects on survival and quantifying the associated uncertainty. Outside the causal survival setting, BART has also been used for survival analysis with discrete-time hazards \citep{sparapani2016nonparametric}, semiparametric survival models \citep{bonato2011bayesian}, and competing-risk and recurrent-event models \citep{sparapani2020competing,sparapani2020recurrent}. Subsequent work has extended these ideas to clustered, interval-censored, and spatial survival settings \citep{basak2022semiparametric,ghosh2024analysis}; fully nonparametric survival models \citep{linero2022bayesian,alam2025unified,sparapani2023nonparametric}; dynamic treatment regimes \citep{li2024dynamic}; and relative survival models \citep{basak2025understanding}.

Several existing causal machine learning methods can accommodate a cured subpopulation. For example, \citet{cui2023estimating} proposed \emph{causal survival forests} (CSFs), which extend causal forests to survival settings and target finite-horizon survival functionals such as survival probabilities and restricted mean survival time. Under the assumption that there exists a time $\tau$ beyond which individuals are considered cured, these methods remain applicable in the presence of a cured subpopulation. Another common approach for modeling survival data with a cured subpopulation is the \emph{mixture cure rate} modeling; \citet{sun2025tree} adopt this framework, using a BART probit model for the probability of being cured and a Bayesian causal forest (BCF, \citealp{hahn2020bayesian}) model on log-survival times among uncured individuals. Unlike the promotion-time models considered here, mixture cure models parameterize the cure probability and the survival distribution among uncured individuals separately. Consequently, these approaches do not explicitly share information between the cure and latency components of the model.

\section{Notation and Definition of Causal Parameters}
\label{sec:definition}

We adopt standard notation from causal survival analysis. Let $T_i$ denote the survival time, $C_i$ a right-censoring time, $X_i$ a vector of confounders and effect modifiers, and $A_i$ a binary treatment indicator. Rather than observing $(T_i, C_i)$ directly, we observe only the event time $Y_i = \min(T_i, C_i)$ and the event indicator $\delta_i = 1(T_i \leq C_i)$. Unlike in standard survival settings, we allow individual $i$ to be \emph{cured}, in which case $T_i = \infty$; such individuals are therefore necessarily censored. We use the potential outcomes framework to define causal effects \citep{rubin2005causal}. For each treatment level $a \in \{0,1\}$, let $T_i(a)$ denote the potential survival time under treatment $a$. Under the consistency assumption \citep{rubin1980randomization}, the realized outcome is $T_i = T_i(A_i)$. We also define the propensity score $e(x) = \Pr(A_i = 1 \mid X_i = x)$.

\paragraph{Causal Effects}
To define the causal effects of interest, it is useful to define some additional potential outcomes. First, we define the \emph{potential outcomes for the survival status indicator}
\begin{math}
  D_i(a, t) = 1\{T_i(a) > t\},
\end{math}
for $t \in [0, \infty]$. We then define the average causal effect on survival status at time $t$ \citep{chen2001causal} and the causal effect on the cure event as
\begin{align*}
  \Delta_S(t) = \E\{D_i(1, t) - D_i(0, t)\}
  \quad \text{and}
  \quad
  \Delta_C = \E\{D_i(1, \infty) - D_i(0,\infty)\},
\end{align*}
respectively. Note that $ \E\{D_i(a, t)\} = \Pr(T_i(a) > t) = S_a(t)$, so $\Delta_S(t)= S_1(t) - S_0(t)$ is the average causal effect on the probability of surviving beyond time $t$. Note also that $D_i(a, \infty)$ is never observed, and $\Delta_C$ therefore is only recoverable under assumptions about the tail behavior of $S_a(t)$. As a measure of lifetime gained up to time $t$ due to treatment, we define the \emph{causal effect on restricted mean survival time (RMST)} \citep{chen2001causal} as
\begin{align*}
  \Delta_R(t) = \E\{R_i(1, t) - R_i(0, t)\}
  \quad \text{where}
  \quad
  R_i(a,t) = \min\{T_i(a), t\}.
\end{align*}
RMST can be interpreted as the life-expectancy-up-to-time-$t$ \citep{royston2013restricted} and so $\Delta_R(t)$ represents the average amount of lifetime gained due to treatment up to time $t$. Alternatively, one can write the causal RMST effect as the integrated difference in survival functions $\Delta_R(t) = \int_0^t [\Pr\{T_i(1) > u\} - \Pr\{T_i(0) > u\}] \ du =  \int_0^t \{S_1(u) - S_0(u)\} \ du$.

Conditional average versions of each of these effects can also be defined
straightforwardly. For example, the conditional average treatment effect (CATE) version of $\Delta_R(t)$ is $\Delta_R(t,x) = \E\{R_i(1, t) - R_i(0, t) \mid X_i = x\}$. Similarly, we can define $\Delta_C(x)$, $\Delta_S(t, x)$, and so forth.

\subsection{Identification Assumptions}

Throughout, we make the standard positivity, ignorability, stable unit treatment value, consistency, and ignorable censoring assumptions that are standard for identifying causal effects in causal survival analysis. For completeness, we list these in the Supplementary Material as Assumptions~1---4. We also make the following assumption to identify the cured subpopulation.

\paragraph{Assumption 5: Cure Threshold.}
There exists a known time $\tau < \infty$ such that the hazard function satisfies $h(t \mid a, x) = 0$ for all $t > \tau$, $a \in \{0,1\}$, and $x$ in the support of $X_i$. Equivalently, survival beyond $\tau$ implies cure. Additionally, we require sufficient follow-up in the sense that $\Pr(C_i > \tau \mid X_i = x) > 0$ for all $x$.

\vspace{1em}

Some tail behavior assumption for the survival function is required in order to conclude that an apparent asymptote in the survival curve truly reflects the presence of a cured subpopulation, since the event $\{T_i = \infty\}$ is never directly observed. Assumption 5, which links $\{T_i = \infty\}$ to the observable event $\{T_i > \tau\}$, is a pragmatic option that is frequently adopted in the semiparametric literature on cure rate models \citep{peng2014cure}. While not strictly required, it has several appealing consequences for finite-sample identifiability and estimation stability, both of which can otherwise be problematic even when the model is formally identified \citep{li2001identifiability}.

Our methodology can also accommodate the weaker assumption that there exists some $\tau > 0$ such that the hazard is constant beyond $\tau$, \textit{i.e.}, $\htt(t \mid a, x) = c(a, x)$ for some $c(\cdot, \cdot)$ and $t > \tau$, which also identifies the causal effects. This replaces the sharp cutoff beyond $\tau$ with an exponential tail. See \citet{peng2014cure} for a discussion of this point. The key requirement for identifiability is that the tail behavior of $[T_i \mid T_i < \infty, A_i, X_i]$ is sufficiently constrained for the model to determine where the survival curves level off.

\subsection{The Latency and Stochastic Latency Effects}
\label{sec:stochastic-intervention-effects}

The effects $\Delta_S(t)$ and $\Delta_R(t)$ are standard causal estimands in survival analysis, whereas $\Delta_C$ is specific to cure rate models. Defining a companion causal effect to $\Delta_C$ that removes the contribution of cure is desirable, but doing so is not straightforward and involves potential pitfalls. For example, the uncured average treatment effect (UATE) proposed by \citet{sun2025tree}, $\Delta_{\text{uncured}}(x) = \E\{\log T_i(1) \mid X_i = x,\, T_i(1) < \infty\} - \E\{\log T_i(0) \mid X_i = x,\, T_i(0) < \infty\}$, does not correspond to a standard causal estimand because it compares potential outcomes across two different latent populations: individuals who would remain uncured under treatment and individuals who would remain uncured under control. \citet{sun2025tree} discuss further assumptions required for $\Delta_{\text{uncured}}$ to admit a causal interpretation.

As a first step in the direction of defining a causal effect that ``removes'' the effect of the cure process from RMST we define the \emph{latency effect}
\begin{align*}
  \Delta_L(t) = \E\{L_i(1, t) - L_i(0, t)\}
  \quad \text{where} \quad
  L_i(a, t) = R_i(a, t) - t \, D_i(a, \infty).
\end{align*}
Equivalently, one can show that
\begin{math}
  \E\{L_i(a, t)\} = \int_0^t \Pr(u < T_i(a) < \infty) \, du,
\end{math}
so that $\Delta_L(t) = \int_0^t \big[\Pr(u < T_i(1) < \infty) - \Pr(u < T_i(0) < \infty)\big] \, du$ represents the integrated difference in the probability of being alive and uncured at time $u$, for $u \in [0, t]$. This implies that $\Delta_L(t)$ measures how much more time, on average up to horizon $t$, a treated individual spends alive \emph{and} uncured relative to an untreated individual.

The value of defining the latency effect is that it allows for decomposition of $\Delta_R(t)$ into a part that is directly attributable to the probability of being cured and a part that is not; in particular, $\Delta_R(t) = \Delta_L(t) + t \, \Delta_C$. Moreover, $\Delta_L(t)$ is a valid causal effect in the sense that it contrasts two well-defined potential outcomes.

Unfortunately, $\Delta_L(t)$ does not have a clean clinical interpretation. We investigate this by introducing the principal strata \citep{frangakis2002principal}
\begin{align*}
  UU &= \{i : T_i(0) < \infty, T_i(1) < \infty\}, &
  UC &= \{i : T_i(0) < \infty, T_i(1) = \infty\}, \\
  CU &= \{i : T_i(0) = \infty, T_i(1) < \infty\}, &
  CC &= \{i : T_i(0) = T_i(1) = \infty\}.
\end{align*}
A common assumption in this setting is that the treatment is beneficial in the sense that the stratum $CU$ is empty, i.e., nobody who died under treatment would have survived under control. Under this assumption, we can express $\Delta_L(t)$ in terms of its contribution to the causal effect on RMST in $UU$. The following proposition relates these two quantities; its main implication is that individuals who are cured by treatment contribute $-\Delta_C \, \E\{R_i(0,t) \mid i \in UC\}$ to $\Delta_L(t)$, which can make the treatment appear harmful (as measured by $\Delta_L(t)$) simply because the treatment is effective at curing individuals.

\begin{proposition}
  \label{prop:uu}
  Let $\Delta_{UU}(t) = \E\{R_i(1, t) - R_i(0,t) \mid i \in UU\}$ and $p_1 = \Pr\{T_i(1) < \infty\}$. Suppose that Assumptions~1---4 hold and that $\Pr(i \in CU \mid X_i) = 0$. Then 
  \begin{align*}
    \Delta_L(t)
    =
    p_1 \Delta_{UU}(t) - \Delta_C \E\{R_i(0,t) \mid i \in UC\}.
  \end{align*}
\end{proposition}

All proofs are deferred to the Supplementary Material. This flaw in $\Delta_L(t)$ motivates us to find an alternate decomposition of $\Delta_R(t)$ that does not have this flaw. Let $p_a(x) = \Pr\{T_i(a) < \infty \mid X_i = x\}$ denote the probability of not being cured under treatment $a$, and define the conditional RMST among uncured individuals as $m_a(t,x) = \E\{R_i(a,t) \mid T_i(a) < \infty, X_i = x\}$. Equivalently, if we let $G_a(\cdot \mid x)$ denote the survival function of $T_i(a)$ conditional on not being cured,
then $m_a(t,x) = \int_0^t G_a(u \mid x) \ du$. With these definitions, we have $\E\{R_i(a,t) \mid X_i = x\} = \{1 - p_a(x)\} \times t + p_a(x) \times m_a(t,x)$.

Next we define a synthetic conditional distribution $Q_{aa'}(\cdot \mid x)$ by combining the cure probability under treatment $a$ with the finite-event-time distribution under $a'$ as $Q_{aa'}(\cdot \mid x) = \{1 - p_a(x)\} \, \delta_\infty(\cdot) + p_{a}(x) \, G_{a'}(\cdot \mid x)$. Let $T_i(a, a') \sim Q_{aa'}(\cdot \mid X_i)$ denote a draw from this distribution conditional on $X_i$ and define $R_i(a,a',t) = \min\{T_i(a,a'), t\}$. The corresponding stochastic-intervention RMST is
\begin{math}
  \vartheta_t(a,a') = \E\{R_i(a,a',t)\} = \E[\{1 - p_a(X_i)\} t + p_a(X_i) m_{a'}(t, X_i)]
\end{math}
and we define the stochastic cure component of the RMST effect as
\begin{align*}
  \Delta_{SC}(t) &= \frac{\vartheta_t(1,0) - \vartheta_t(0,0)}{2} + \frac{\vartheta_t(1,1) - \vartheta_t(0,1)}{2} \\
  &=
  \E\left[ \{p_0(X_i) - p_1(X_i)\} \left\{ t - \frac{m_0(t,X_i) + m_1(t,X_i)}{2} \right\} \right].
\end{align*}
The intuition behind this definition is that we imagine that our intervention can separately be used to replace $p_1(x)$ with $p_0(x)$ (i.e., modifying the effect of the cure) and to replace $G_1(\cdot \mid x)$ with $G_0(\cdot \mid x)$ (i.e., modifying the effect on survival among those who are uncured). It is stochastic in the sense that the effect can be expressed as $\E\{\vartheta_t(1, A') - \vartheta_t(0, A')\}$ where $A' \sim \Bernoulli(1/2)$, so that we randomly choose which survival distribution we assign.

The stochastic cure effect has a complementary latency component, defined as
\begin{align*}
  \Delta_{SL}(t) = 
  \E\left[ \{m_1(t, X_i) - m_0(t, X_i)\} \left\{\frac{p_0(X_i) + p_1(X_i)}{2} \right\} \right].
\end{align*}
Similarly, it can be expressed as $\Delta_{SL}(t) = \E\{\vartheta_t(A',1) - \vartheta_t(A',0)\}$, where $A' \sim \Bernoulli(1/2)$. 
We write $\Delta_{SC}(t,x)$ and $\Delta_{SL}(t,x)$ for the corresponding conditional quantities.
As these effects are functionals of the observed data distribution, both are identified under Assumptions~1---5. Moreover, they decompose the RMST effect.

\begin{proposition}
  \label{prop:stochastic-decomposition}
  For every time horizon $t$ and covariate value $x$,
  $\Delta_R(t,x) = \Delta_{SC}(t,x) + \Delta_{SL}(t,x)$.
\end{proposition}


The proposed stochastic intervention effects attempt to address the main interpretational difficulty of $\Delta_L(t)$, which implicitly attributes the entire $t$ units of RMST in a cured individual to the effect of cure. By contrast, $\Delta_{SC}(t)$ treats the gain from cure as the difference between $t$ and a reference finite-event-time RMST. Thus, if treatment increases the cure probability but the finite-event-time distribution is otherwise unchanged, $\Delta_{SC}(t)$ captures the full RMST effect, while conversely if the treatment does not change the cure probability then $\Delta_{SC}(t) = 0$ and the entire RMST effect is attributed to changes in the finite-event-time distribution.

Without further assumptions, we view $\Delta_{SC}(t)$ and $\Delta_{SL}(t)$ as model-based standardizations of the counterfactual survival distributions rather than as literal mechanistic effects. A mechanistic interpretation would require one to imagine treatment components that can separately alter cure incidence and post-susceptibility latency, bringing the estimands close in spirit to separable effects or interventional mediation effects \citep{stensrud2022separable,vanderweele2014effect}.
To link the stochastic effects to more mechanistic effects, $\Delta_{SL}(t)$ can also be related to effects in the principal strata.

\begin{proposition}
  \label{prop:uu2}
  Suppose that Assumptions~1---4 hold and that $\Pr(i \in CU \mid X_i = x) = 0$.
  Let $p_a(x) = \Pr\{T_i(a) < \infty \mid X_i = x\}$ and suppose $p_0(x) > 0$.
  Let $\Delta_{s}(t,x) = \E\{R_i(1,t) - R_i(0,t) \mid i \in s, X_i = x\}$
  and $B_s(t,x) = \E\{R_i(0,t) \mid i \in s, X_i=x\}$ for $s \in \{UU,UC\}$.
  Then
\begin{align*}
    \Delta_{SL}(t,x)
    &=
      \left\{
      \frac{p_1(x) + p_0(x)}{2}
      \right\}
      \left\{
    \Delta_{UU}(t,x)
    +
    \frac{\Delta_C(x) \{B_{UU}(t,x) - B_{UC}(t,x)\}}{p_0(x)}
      \right\}
    .
  \end{align*}
  Hence $\Delta_{SL}(t,x)$ is proportional to $\Delta_{UU}(t,x)$ if $\Delta_C(x) = 0$ or  $B_{UU}(t,x) = B_{UC}(t,x)$.
\end{proposition}

Relative to Proposition~\ref{prop:uu}, $\Delta_{SL}(t,x)$ is, informally, a ``doubly-robust'' proxy for $\Delta_{UU}(t,x)$: it can be proportional to $\Delta_{UU}(t,x)$ by either (i) having $\Delta_C(x) = 0$ (i.e., there is no cure effect) or (ii) by satisfying $B_{UU}(t,x) = B_{UC}(t,x)$. The latter is a principal ignorability condition stating that, for individuals who are uncured under control, knowing whether they would be cured under treatment does not inform their RMST under control. This type of principal ignorability is discussed by \citet{wang2024causal}. Consequently, the relationship in Proposition~\ref{prop:uu2} is much more desirable than the relationship in Proposition~\ref{prop:uu}. Proposition~\ref{prop:uu2} might also form the basis of a sensitivity analysis by varying the sensitivity parameter $\rho(x) = B_{UU}(t,x) - B_{UC}(t,x)$ over some plausible range of values, although we do not pursue this.

\paragraph{Share of RMST Attributable to Cure}
A useful summary of the overall effect of the cure probability is the signed cure share $\Delta_{SC}(t) / \Delta_R(t)$, provided that $\Delta_R(t)$ is not close to zero. Because the two components may have opposite signs, this ratio need not lie between zero and one. If a bounded descriptive measure is desired, one can instead report $|\Delta_{SC}(t)| / \{|\Delta_{SC}(t)| + |\Delta_{SL}(t)|\}$, which can be interpreted as the relative magnitude of the cure component rather than as a percentage of the net effect.

\section{Bayesian Machine Learning for Cure Rate Models}


To estimate the causal effects described in Section~\ref{sec:definition}, we propose machine learning methods based on the promotion-time cure model \citep{tsodikov2003estimating}
\begin{align}
  \label{eq:promotion-time}
  S(t \mid a, x) = \exp\left\{ -\theta(a,x) \, \Ft(t \mid a,x) \right\},
\end{align}
where $\Ft(t \mid a, x)$ is a cumulative distribution function. Because $\lim_{t \to \infty} \Ft(t \mid a, x) = 1$, this implies that the cure probability is given by
\begin{math}
  \lim_{t \to \infty} S(t \mid a, x) = e^{-\theta(a,x)}.
\end{math}
The form of the promotion-time model is motivated by assuming a latent number of competing risks $N_i(a) \sim \text{Poisson}(\theta(a,x))$ for individual $i$ under treatment $a$. If $N_i(a) = 0$, the individual is cured and $T_i(a) = \infty$; otherwise, each of the $N_i(a)$ latent failure times are drawn \textit{i.i.d.}~from the distribution with distribution function  $\Ft(t \mid a, x)$, and the survival time $T_i(a)$ is the minimum of these latent failure times (see \citealp{chen1999new} for a derivation). 

An advantage of the promotion-time model is that it more effectively borrows information across the cure probability $e^{-\theta(a,x)}$ and the survival function conditional on not being cured
\begin{math}
  G_a(t \mid x) = \{e^{-\theta(a,x) \, \Ft(t \mid a, x)} - e^{-\theta(a,x)}\} / \{1 - e^{-\theta(a,x)}\}
\end{math},
because both quantities are determined by the same underlying functions
$\theta(a,x)$ and $\Ft(t \mid a,x)$.
By contrast, nonparametric mixture cure models such as the one used by \citet{sun2025tree} parameterize the mixture probability separately from the survival distribution, and consequently do not share information across these components.


\subsection{A Brief Review of BART}

To model the unknown functions $\theta(a,x)$ and $\Ft(t \mid a, x)$ we use \emph{Bayesian additive regression trees} (BART) models. A general BART model expresses a function of interest as a sum of regression trees
\begin{math}
  r(a, x) = \sum_{m=1}^M g(a, x; \Tree_m, \sM_m),
\end{math}
where $\Tree_m$ denotes the tree shape and splitting rules of tree $m$ and $\sM_m = \{\mu_{\ell m} : \ell \in \Leaves(\Tree_m)\}$ consists of the leaf node predictions such that $g(a, x ; \Tree_m, \sM_m) = \mu_{\ell m}$ if $(a, x)$ is associated to leaf node $\ell$ of $\Tree_m$. Each $\Tree_m$ is a binary tree whose internal (non-terminal) nodes are equipped with \emph{splitting rules} of the form $[x_j \le c]$, where $x_j$ is a component of the input vector $(a, x)$ and $c$ is a cutpoint. An input arriving at an internal node is associated to the left child if the splitting condition is satisfied, and to the right child otherwise, until a terminal (leaf) node is reached.

BART was introduced by \citet{chipman2010bart} for nonparametric regression with normal errors, as well as probit regression. In causal inference, \citet{hill2011bayesian} used BART to estimate heterogeneous treatment effects in observational studies, with the main goal of flexibly adjusting for confounding and recovering individualized counterfactual outcome surfaces without having to prespecify interactions or nonlinear main effects. Extending this, \citet{hahn2020bayesian} proposed the \emph{Bayesian causal forest} (BCF), which reparameterizes the outcome model into separate prognostic and treatment-effect components so that treatment effect heterogeneity can be regularized more aggressively.

\subsection{The One-Forest and Two-Forest Models}

We now describe the \BartCure\ model for estimation of the causal estimands defined in Section~\ref{sec:definition}. We introduce two variants of \BartCure, which we refer to as the \emph{one-forest} and \emph{two-forest} models, respectively. These models are fully nonparametric, in the sense that they can approximate any conditional survival function $S(t \mid a, x)$. The two-forest approach directly applies the BART model of \citet{chipman2010bart} to \eqref{eq:promotion-time} by setting
\begin{align*}
  \log \theta(a,x) = \sum_{m = 1}^M g(a, x; \Tree^\theta_m, \sM^\theta_m)
  \ \ \text{and} \ \
  \htt(t \mid a,x) = 
  \lambda_0(t) \exp\left\{ \sum_{m = 1}^M g\big(B(t), a, x; \Tree^h_m, \sM^h_m\big) \right\},
\end{align*}
where $B(\cdot): \Reals^+ \to \{1, \ldots, K\}$ maps a given time $t$ to a
\emph{bin} in which the hazard function is constant, $\lambda_0(t)$ is a baseline hazard function, and $\htt(t \mid a, x) = -\frac{d}{dt} \log \{1 - \Ft(t \mid a, x)\}$ is the hazard function associated to $\Ft(t \mid a, x)$. We adopt the shorthand $r^\theta(a,x) = \log \theta(a,x)$ and $r^h(t, a, x) = \sum_m g\big(B(t), a, x; \Tree_m^h, \sM_m^h \big)$.

The one-forest model instead reparameterizes the model as
\begin{align}
  \label{eq:one-forest}
  \theta(a,x) \, \ft(t \mid a, x) = \lambda_0(t) e^{r(t, a, x)}
  \quad \text{where} \quad
  r(t, a, x) = \sum_{m = 1}^M g\big( B(t), a, x ; \Tree_m, \sM_m  \big).
\end{align}
and where $\ft(t \mid a, x) = \frac{\partial}{\partial t} \Ft(t \mid a, x)$ is the conditional density of $\Ft(t \mid a, x)$. From this, we can recover $\theta(a,x) = H(\infty \mid a, x)$ and $\Ft(t \mid a, x) = H(t \mid a, x) / H(\infty \mid a, x)$ where $H(t \mid a, x) = \int_0^t \lambda_0(u) e^{r(u,a,x)} \ du$. The one-forest model is less faithful to the promotion-time parameterization, but retains its important benefits: it is computationally simple, and allows the cure probability to be a priori correlated with better survival prognosis. A potential drawback of the one-forest model is that sharing information between $\theta(a,x)$ and $\ft(t \mid a, x)$ may not always be desirable.

\paragraph{Prior on $\Tree_m$}
All of the tree structures are assigned a branching process prior as described by \citet{chipman2010bart,chipman1998bayesian}. Each node at depth $d$ is non-terminal with probability
\begin{math}
  \frac{\gamma}{(1 + d)^{\beta}},
\end{math}
where $\gamma \in (0, 1)$ and $\beta > 0$ control the tree size. The default values $\gamma = 0.95$ and $\beta = 2$ encourage trees with few terminal nodes, so that each tree captures only a small piece of the overall signal. Because treatment effect heterogenity only arises through splitting on $a$ and one other variable, this prior naturally penalizes treatment effect heterogeneity.

\paragraph{Prior on $\sM_m$}
To facilitate computations, log-gamma priors are specified for the leaf node predictions $\mu_{\ell m } \sim \log \Gam(a_\mu, b_\mu)$. In this setting, as well as other survival settings, the log-gamma prior is conditionally conjugate \citep{basak2025understanding,linero2022bayesian,alam2025unified}. Following \citet{basak2025understanding}, we choose $a_\mu$ and $b_\mu$ so that $\E(\mu_{\ell m}) = 0$ and $\Var(\mu_{\ell m}) = \sigma^2_\mu / M$ with respect to the prior, with $\sigma_\mu = 1.5$ in all of our examples.

\paragraph{Prior for $\lambda_0(t)$}
For both the one-forest and two-forest models, we model the baseline hazard $\lambda_0(t)$ as piecewise-constant with cut points $0 = \alpha_0 < \alpha_1 < \cdots < \alpha_{K - 1} < \alpha_K = \infty$ such that $\lambda_0(t) = \sum_{k = 1}^K \lambda_k \times 1(\alpha_{k-1} \le t < \alpha_k)$. We impose Assumption~5 by setting $\alpha_{K -1} = \tau$ and $\lambda_K = 0$, although this is not strictly required to fit the two-forest model. We then set $\lambda_k \iid \Gam(a_\lambda, b_\lambda)$ with $a_\lambda = 1$ and $b_\lambda \sim \Gam(1, 1)$ by default.

\paragraph{Targeted Selection}
In observational studies, treatment assignment may depend strongly on the baseline prognosis, a phenomenon referred to as \emph{targeted selection} \citep{hahn2020bayesian}. To mitigate the impact of targeted selection, it is recommended to include an estimate of the propensity score as a covariate in the model. We follow this approach by incorporating an estimated propensity score into both the one-forest and two-forest \BartCure\ models.

\paragraph{Computation}
Inference for both the one-forest and two-forest models is performed via Gibbs sampling using the generalized Bayesian backfitting algorithm of \citet{hill2020bayesian}. In the case of the one-forest model, the algorithm is essentially identical to the algorithm of \citet{alam2025unified} except that the cure threshold is taken into account. A Gibbs sampler for the two-forest model is not immediate from prior work, but is feasible after augmenting the latent number of unobserved events $N_i \sim \Poisson\{\theta(A_i, X_i) \, [1 - \Ft(Y_i \mid A_i, X_i)]\}$. Importantly for computational feasibility, the latent events themselves do not need to be sampled. Details are deferred to the Supplementary Material.

\section{Experiments}
\label{sec:experiments}

We now conduct simulation experiments to evaluate the performance of the one-forest and two-forest \BartCure\ models relative to baseline methods. Our goal is to determine whether \BartCure\ provides reliable estimation of average causal effects, conditional average causal effects, and treatment effect heterogeneity, in the presence of a cured subpopulation. 

\subsection{Estimating Causal Effects}

We compare \BartCure\ to competing methods on a suite of data generating processes (DGPs) that were introduced in prior work on causal survival analysis, augmented (or not) to have a cured subpopulation.

\paragraph{Methods for Comparison} 
We compare the one-forest model and two-forest models, which we label \BartCure(1) and \BartCure(2). We also consider two existing causal machine learning survival analysis methods: causal survival forests (CSF, \citealp{cui2023estimating}) and an accelerated failure time BART model (IndivAFT, \citealp{henderson2020individualized}). CSF is implemented in the \texttt{grf} package \citep{tibshirani2024grf}, while IndivAFT is implemented in the \texttt{AFTrees} package. IndivAFT sets $\log T_i(a) = \mu_a(X_i) + \epsilon_i$ with a flexible Dirichlet process mixture model specified for the error distribution.
To address targeted selection \citep{hahn2020bayesian} all BART methods incorporated the estimate propensity score as a covariate, while the CSF uses the propensity score via inverse propensity score weighting. All BART methods use 200 trees and inferences are based on 2,000 draws from a Gibbs sampler targeting the posterior.

\paragraph{Data Generating Processes}
We consider three families of DGPs drawn from the causal survival analysis literature \citep{henderson2020individualized,cui2023estimating,hu2021estimating}. The same DGPs were analyzed by \citet{kabata2025quantifying}; full details on these DGPs are in the Supplementary Material. All experiments used $N = 1000$ samples, which is somewhat smaller than the sample size of the CALGB 40101 breast cancer trial.
Each of these DGPs was extended to include a cured fraction via the mixture cure model
\begin{math}
  S(t \mid a, x)
  = \{1 - p_a(x)\} + p_a(x) G_a(t \mid x)
\end{math}
where $G_a(t \mid x)$ is the survival function of the uncured sub-population, obtained by truncating the base survival at the time horizon $t$. We set the cure probability to $\expit\{\alpha + z_a(x)\}$ where $z_a(x)$ is $\logit G_a(t \mid x)$ standardized to have mean $0$ and variance $1$, and $\alpha$ is chosen so that the marginal cure rate is $1/2$ under all settings. This implies that individuals with a high probability of surviving until the follow-up horizon $t$ in the uncured population also have a high probability of being cured.

\paragraph{Comparison Measures}
For the simulation settings without a cured subpopulation, we estimate $\Delta_R(t), \Delta_R(t, x)$, $\Delta_S(t)$, and $\Delta_S(t, x)$ at the fixed time horizon $t$. For settings with a cured subpopulation, we monitor $\Delta_C$ and $\Delta_C(x)$ rather than $\Delta_S(t)$ and $\Delta_S(t,x)$. We then compute the bias, root mean-squared error (RMSE), and the length and coverage of nominal 95\% confidence/credible intervals, averaged over all simulated datasets and individuals.

\subsection{Results for Estimating Causal Effects}

Table~\ref{tab:ate-cure} and Table~\ref{tab:ate-nocure} report the results for the causal effects in settings with and without a cured subpopulation, respectively. We defer simulation results for conditional average treatment effects to the Supplementary Material, with results in Figures~\ref{fig:cure_cui}---\ref{fig:sim_hu}.

\begin{table}
\centering
\scalebox{0.79}{
\begin{tabular}[t]{lrlrrrrrrrr}
\toprule
\multicolumn{3}{c}{ } & \multicolumn{4}{c}{$\Delta_C$} & \multicolumn{4}{c}{$\Delta_R(t)$} \\
\cmidrule(l{3pt}r{3pt}){4-7} \cmidrule(l{3pt}r{3pt}){8-11}
Setting & DGP & Method & Bias & RMSE & Coverage & CI Length & Bias & RMSE & Coverage & CI Length\\
\midrule
\multirow{8}{*}{Cui} & \multirow{4}{*}{1} & \BartCure(1) & -0.014 & 0.029 & 0.95 & 0.108 & -0.012 & 0.024 & 0.94 & 0.090\\
 & & \BartCure(2) & -0.011 & 0.040 & 0.99 & 0.183 & -0.009 & 0.023 & 0.95 & 0.090\\
 & & CSF & 0.005 & 0.030 & 0.94 & 0.119 & 0.003 & 0.022 & 0.96 & 0.092\\
 & & IndivAFT & -0.011 & 0.026 & 0.94 & 0.087 & -0.008 & 0.021 & 0.95 & 0.076\\
\cmidrule(l{3pt}){2-11}
 & \multirow{4}{*}{2} & \BartCure(1) & -0.010 & 0.031 & 0.94 & 0.117 & -0.003 & 0.034 & 0.96 & 0.134\\
 & & \BartCure(2) & 0.005 & 0.044 & 0.96 & 0.190 & 0.000 & 0.034 & 0.96 & 0.133\\
 & & CSF & 0.000 & 0.034 & 0.95 & 0.131 & -0.002 & 0.035 & 0.95 & 0.139\\
 & & IndivAFT & 0.015 & 0.031 & 0.91 & 0.109 & 0.022 & 0.039 & 0.91 & 0.130\\
\midrule
\multirow{16}{*}{Henderson} & \multirow{4}{*}{1} & \BartCure(1) & -0.006 & 0.028 & 0.93 & 0.109 & -0.010 & 0.100 & 0.95 & 0.401\\
 & & \BartCure(2) & -0.013 & 0.041 & 1.00 & 0.217 & -0.011 & 0.104 & 0.94 & 0.413\\
 & & CSF & 0.006 & 0.030 & 0.97 & 0.123 & 0.014 & 0.105 & 0.98 & 0.460\\
 & & IndivAFT & -0.008 & 0.025 & 0.94 & 0.096 & -0.008 & 0.095 & 0.97 & 0.408\\
\cmidrule(l{3pt}){2-11}
 & \multirow{4}{*}{2} & \BartCure(1) & -0.005 & 0.028 & 0.95 & 0.108 & -0.012 & 0.098 & 0.95 & 0.383\\
 & & \BartCure(2) & -0.016 & 0.043 & 1.00 & 0.218 & -0.011 & 0.100 & 0.95 & 0.393\\
 & & CSF & 0.006 & 0.031 & 0.95 & 0.123 & 0.015 & 0.098 & 0.99 & 0.433\\
 & & IndivAFT & -0.020 & 0.028 & 0.59 & 0.056 & -0.047 & 0.094 & 0.91 & 0.320\\
\cmidrule(l{3pt}){2-11}
 & \multirow{4}{*}{3} & \BartCure(1) & -0.005 & 0.028 & 0.94 & 0.109 & -0.013 & 0.099 & 0.94 & 0.389\\
 & & \BartCure(2) & -0.018 & 0.043 & 0.99 & 0.214 & -0.011 & 0.101 & 0.95 & 0.397\\
 & & CSF & 0.006 & 0.030 & 0.95 & 0.123 & 0.016 & 0.099 & 0.99 & 0.438\\
 & & IndivAFT & -0.022 & 0.029 & 0.54 & 0.052 & -0.053 & 0.098 & 0.89 & 0.312\\
\cmidrule(l{3pt}){2-11}
 & \multirow{4}{*}{4} & \BartCure(1) & -0.005 & 0.029 & 0.94 & 0.108 & -0.008 & 0.084 & 0.96 & 0.321\\
 & & \BartCure(2) & -0.017 & 0.042 & 1.00 & 0.212 & -0.007 & 0.086 & 0.97 & 0.336\\
 & & CSF & 0.006 & 0.031 & 0.96 & 0.123 & 0.008 & 0.091 & 0.98 & 0.386\\
 & & IndivAFT & -0.008 & 0.026 & 0.93 & 0.099 & -0.001 & 0.090 & 0.95 & 0.362\\
\midrule
\multirow{16}{*}{Hu} & \multirow{4}{*}{1} & \BartCure(1) & 0.013 & 0.033 & 0.93 & 0.123 & 0.000 & 0.001 & 0.96 & 0.004\\
 & & \BartCure(2) & 0.061 & 0.075 & 0.89 & 0.251 & 0.000 & 0.001 & 0.97 & 0.004\\
 & & CSF & -0.021 & 0.040 & 0.88 & 0.122 & -0.001 & 0.001 & 0.88 & 0.004\\
 & & IndivAFT & 0.069 & 0.070 & 0.00 & 0.054 & 0.002 & 0.002 & 0.36 & 0.003\\
\cmidrule(l{3pt}){2-11}
 & \multirow{4}{*}{2} & \BartCure(1) & -0.006 & 0.033 & 0.96 & 0.122 & 0.000 & 0.001 & 0.95 & 0.004\\
 & & \BartCure(2) & 0.058 & 0.078 & 0.77 & 0.248 & 0.000 & 0.001 & 0.95 & 0.004\\
 & & CSF & -0.014 & 0.039 & 0.90 & 0.121 & 0.000 & 0.001 & 0.92 & 0.004\\
 & & IndivAFT & 0.101 & 0.101 & 0.00 & 0.053 & 0.003 & 0.003 & 0.00 & 0.003\\
\cmidrule(l{3pt}){2-11}
 & \multirow{4}{*}{3} & \BartCure(1) & -0.009 & 0.031 & 0.96 & 0.118 & 0.001 & 0.001 & 0.80 & 0.004\\
 & & \BartCure(2) & 0.130 & 0.146 & 0.54 & 0.300 & 0.001 & 0.001 & 0.95 & 0.004\\
 & & CSF & -0.008 & 0.032 & 0.91 & 0.114 & 0.000 & 0.001 & 0.90 & 0.004\\
 & & IndivAFT & 0.334 & 0.356 & 0.00 & 0.077 & 0.012 & 0.013 & 0.01 & 0.004\\
\cmidrule(l{3pt}){2-11}
 & \multirow{4}{*}{4} & \BartCure(1) & -0.004 & 0.032 & 0.92 & 0.119 & 0.001 & 0.001 & 0.80 & 0.004\\
 & & \BartCure(2) & 0.197 & 0.213 & 0.25 & 0.336 & 0.001 & 0.001 & 0.89 & 0.004\\
 & & CSF & -0.004 & 0.034 & 0.88 & 0.114 & 0.000 & 0.001 & 0.92 & 0.004\\
 & & IndivAFT & 0.304 & 0.351 & 0.00 & 0.078 & 0.010 & 0.012 & 0.01 & 0.004\\
\bottomrule
\end{tabular}
}

\caption{Simulation results for average treatment effects with a cured fraction.}
\label{tab:ate-cure}
\end{table}

\begin{table}
\centering
\scalebox{0.79}{
\begin{tabular}[t]{lrlrrrrrrrr}
\toprule
\multicolumn{3}{c}{ } & \multicolumn{4}{c}{$\Delta_S(t)$} & \multicolumn{4}{c}{$\Delta_R(t)$} \\
\cmidrule(l{3pt}r{3pt}){4-7} \cmidrule(l{3pt}r{3pt}){8-11}
Setting & DGP & Method & Bias & RMSE & Coverage & CI Length & Bias & RMSE & Coverage & CI Length\\
\midrule
\multirow{8}{*}{Cui} & \multirow{4}{*}{1} & \BartCure(1) & -0.004 & 0.011 & 0.94 & 0.045 & -0.015 & 0.028 & 0.91 & 0.094\\
 & & \BartCure(2) & 0.002 & 0.017 & 0.88 & 0.054 & -0.006 & 0.027 & 0.88 & 0.094\\
 & & CSF & -0.004 & 0.011 & 0.92 & 0.045 & -0.002 & 0.026 & 0.94 & 0.103\\
 & & IndivAFT & -0.006 & 0.011 & 0.89 & 0.035 & -0.012 & 0.026 & 0.88 & 0.090\\
\cmidrule(l{3pt}){2-11}
 & \multirow{4}{*}{2} & \BartCure(1) & -0.010 & 0.019 & 0.93 & 0.065 & -0.012 & 0.043 & 0.93 & 0.164\\
 & & \BartCure(2) & -0.002 & 0.023 & 0.89 & 0.076 & -0.001 & 0.045 & 0.93 & 0.162\\
 & & CSF & 0.014 & 0.028 & 0.93 & 0.093 & 0.002 & 0.045 & 0.94 & 0.174\\
 & & IndivAFT & -0.009 & 0.018 & 0.95 & 0.064 & 0.000 & 0.039 & 0.98 & 0.163\\
\midrule
\multirow{16}{*}{Henderson} & \multirow{4}{*}{1} & \BartCure(1) & -0.002 & 0.023 & 0.97 & 0.101 & -0.029 & 0.191 & 0.97 & 0.794\\
 & & \BartCure(2) & -0.004 & 0.026 & 0.94 & 0.105 & -0.021 & 0.209 & 0.92 & 0.787\\
 & & CSF & -0.045 & 0.398 & 1.00 & 1.354 & -0.218 & 2.521 & 1.00 & 7.964\\
 & & IndivAFT & -0.001 & 0.021 & 0.93 & 0.085 & -0.010 & 0.203 & 0.93 & 0.795\\
\cmidrule(l{3pt}){2-11}
 & \multirow{4}{*}{2} & \BartCure(1) & 0.004 & 0.022 & 0.97 & 0.092 & 0.007 & 0.190 & 0.95 & 0.760\\
 & & \BartCure(2) & 0.004 & 0.026 & 0.94 & 0.098 & 0.009 & 0.210 & 0.93 & 0.748\\
 & & CSF & -0.001 & 0.320 & 1.00 & 1.253 & -0.081 & 2.187 & 1.00 & 8.354\\
 & & IndivAFT & 0.003 & 0.014 & 0.96 & 0.059 & 0.016 & 0.148 & 0.95 & 0.651\\
\cmidrule(l{3pt}){2-11}
 & \multirow{4}{*}{3} & \BartCure(1) & 0.001 & 0.022 & 0.97 & 0.094 & -0.015 & 0.188 & 0.97 & 0.769\\
 & & \BartCure(2) & -0.001 & 0.027 & 0.95 & 0.098 & -0.012 & 0.204 & 0.94 & 0.755\\
 & & CSF & 0.011 & 0.310 & 1.00 & 1.175 & 0.266 & 2.108 & 1.00 & 7.340\\
 & & IndivAFT & 0.000 & 0.015 & 0.95 & 0.061 & -0.011 & 0.159 & 0.95 & 0.638\\
\cmidrule(l{3pt}){2-11}
 & \multirow{4}{*}{4} & \BartCure(1) & 0.001 & 0.019 & 0.95 & 0.078 & -0.029 & 0.171 & 0.94 & 0.657\\
 & & \BartCure(2) & -0.002 & 0.025 & 0.91 & 0.087 & -0.020 & 0.182 & 0.91 & 0.648\\
 & & CSF & -0.013 & 0.262 & 1.00 & 0.949 & -0.005 & 1.452 & 1.00 & 5.175\\
 & & IndivAFT & 0.005 & 0.018 & 0.96 & 0.071 & 0.006 & 0.173 & 0.97 & 0.704\\
\midrule
\multirow{16}{*}{Hu} & \multirow{4}{*}{1} & \BartCure(1) & -0.001 & 0.001 & 0.65 & 0.003 & 0.001 & 0.001 & 0.91 & 0.004\\
 & & \BartCure(2) & -0.003 & 0.003 & 0.75 & 0.010 & -0.002 & 0.003 & 0.75 & 0.006\\
 & & CSF & 0.001 & 0.001 & 0.19 & 0.000 & 0.000 & 0.001 & 0.93 & 0.005\\
 & & IndivAFT & 0.000 & 0.001 & 0.90 & 0.002 & 0.001 & 0.001 & 0.93 & 0.005\\
\cmidrule(l{3pt}){2-11}
 & \multirow{4}{*}{2} & \BartCure(1) & -0.001 & 0.001 & 0.96 & 0.004 & 0.000 & 0.001 & 0.96 & 0.004\\
 & & \BartCure(2) & -0.001 & 0.003 & 0.95 & 0.012 & 0.000 & 0.001 & 0.96 & 0.004\\
 & & CSF & 0.000 & 0.000 & 0.79 & 0.000 & 0.000 & 0.001 & 0.96 & 0.004\\
 & & IndivAFT & -0.001 & 0.002 & 0.85 & 0.005 & 0.000 & 0.001 & 0.96 & 0.004\\
\cmidrule(l{3pt}){2-11}
 & \multirow{4}{*}{3} & \BartCure(1) & -0.003 & 0.003 & 0.22 & 0.006 & 0.000 & 0.001 & 0.95 & 0.003\\
 & & \BartCure(2) & -0.007 & 0.008 & 0.19 & 0.013 & -0.001 & 0.002 & 0.71 & 0.004\\
 & & CSF & 0.001 & 0.001 & 0.20 & 0.000 & 0.000 & 0.001 & 0.93 & 0.004\\
 & & IndivAFT & -0.005 & 0.006 & 0.00 & 0.007 & 0.000 & 0.001 & 0.94 & 0.003\\
\cmidrule(l{3pt}){2-11}
 & \multirow{4}{*}{4} & \BartCure(1) & -0.002 & 0.003 & 0.31 & 0.006 & 0.000 & 0.001 & 0.96 & 0.003\\
 & & \BartCure(2) & -0.007 & 0.007 & 0.29 & 0.013 & -0.002 & 0.002 & 0.71 & 0.004\\
 & & CSF & 0.001 & 0.001 & 0.25 & 0.000 & 0.000 & 0.001 & 0.95 & 0.004\\
 & & IndivAFT & -0.004 & 0.004 & 0.00 & 0.006 & 0.000 & 0.001 & 0.96 & 0.004\\
\bottomrule
\end{tabular}
}

\caption{Simulation results for average treatment effects without a cured fraction.}
\label{tab:ate-nocure}
\end{table}

\paragraph{Average Treatment Effects: With a Cured Subpopulation}
Across all settings, methods perform similarly for estimating average treatment effects across all metrics, with the exception of \BartCure(2), which appears less stable specifically for estimating $\Delta_C$ and so has wider intervals. Aside from this, both IndivAFT and \BartCure(2)\ perform quite poorly on the Hu~3 and Hu~4 DGPs for estimating $\Delta_C$, and IndivAFT performs poorly for $\Delta_R(t)$ as well; in the case of IndivAFT, this occurs because the model does not account for the cured subpopulation, and it also performs poorly for estimating $\Delta_C$ on the Henderson~2 setting. In view of this, the best performing methods for estimating ATEs are \BartCure(1) and the CSF; \BartCure(1)\ performs marginally better by RMSE across most settings and better by coverage on the Hu DGPs, and so performs best overall.

\paragraph{Average Treatment Effects: Without a Cured Subpopulation}
In the absence of a cured subpopulation, the basic findings are the same, with the exception that IndivAFT no longer performs poorly on any of the settings. We see that it is extremely difficult for any method to estimate $\Delta_S(t)$ under the Hu~3 and Hu~4 settings. This occurs because the survival functions are very close to zero at these follow-up times, and so small amounts of bias in estimating $\Delta_S(t)$ dominate inference; coverage for $\Delta_R(t)$, which is determined by the full survival curve, is much better. Interestingly, CSF suffered from substantial instability across all Henderson settings due to its use of inverse weighting highlighting that the BART-based methods can be much more stable.

\subsection{Directionality of Heterogeneity}
\label{sec:directionality}

To complement our results on estimation of individual effects, we also give an assessment of how well the one-forest \BartCure\ model and CSFs perform on the task of estimating the \emph{directionality} of an effect. To test for the existence of individual differences in effect, we constructed for each individual an 80\% confidence/credible interval and checked whether this interval contains $0$. For each individual, we then computed the centered treatment effect $\Delta_R(\tau, x) - \Delta_R(\tau)$ so that positive values indicate above-average benefit. We recorded the \emph{discovery rate}, the \emph{correct sign rate}, the \emph{Type-S error rate}, and the \emph{net directional score},
\begin{align*}
  \DR = \frac{N_D}{N},
  \quad
  \CS = \frac{N_{CD}}{N},
  \quad
  \text{Type-S} = \frac{N_{ID}}{N_D} 1(N_D \ne 0),
  \quad
  \NDS = \frac{N_{CD} - N_{ID}}{N}
\end{align*}
where $N$ is the sample size, $N_D$ is the number of discoveries, $N_{CD}$ is the number of correct discoveries, and $N_{ID}$ is the number of incorrect discoveries. A method that discovers few effects will have a low correct sign rate but also a low Type-S error rate, while a method that declares many effects aggressively may achieve a high correct sign rate at the cost of elevated Type-S errors; the net directional score summarizes the tradeoff. Higher values of the net directional score indicate that a method is both willing and able to recover the direction of treatment-effect heterogeneity.

\begin{table}[t]
\centering
\scalebox{0.84}{
\begin{tabular}[t]{lrlrrrr}
\toprule
Setting & DGP & Method & Discovery rate & Correct sign rate & Type-S error & Net directional score\\
\midrule
\multirow{4}{*}{Cui} & \multirow{2}{*}{1} & \BartCure(1) & 0.021 & 0.010 & 0.531 & -0.001\\
 & & CSF & 0.023 & 0.018 & 0.227 & 0.013\\
\cmidrule(l{3pt}){2-7}
 & \multirow{2}{*}{2} & \BartCure(1) & 0.263 & 0.261 & 0.008 & 0.259\\
 & & CSF & 0.229 & 0.225 & 0.016 & 0.221\\
\midrule
\multirow{8}{*}{Henderson} & \multirow{2}{*}{1} & \BartCure(1) & 0.345 & 0.340 & 0.015 & 0.335\\
 & & CSF & 0.049 & 0.045 & 0.095 & 0.040\\
\cmidrule(l{3pt}){2-7}
 & \multirow{2}{*}{2} & \BartCure(1) & 0.372 & 0.367 & 0.014 & 0.361\\
 & & CSF & 0.072 & 0.068 & 0.063 & 0.063\\
\cmidrule(l{3pt}){2-7}
 & \multirow{2}{*}{3} & \BartCure(1) & 0.372 & 0.367 & 0.013 & 0.362\\
 & & CSF & 0.072 & 0.067 & 0.059 & 0.063\\
\cmidrule(l{3pt}){2-7}
 & \multirow{2}{*}{4} & \BartCure(1) & 0.368 & 0.363 & 0.013 & 0.358\\
 & & CSF & 0.050 & 0.045 & 0.091 & 0.041\\
\midrule
\multirow{8}{*}{Hu} & \multirow{2}{*}{1} & \BartCure(1) & 0.454 & 0.443 & 0.024 & 0.432\\
 & & CSF & 0.537 & 0.511 & 0.048 & 0.485\\
\cmidrule(l{3pt}){2-7}
 & \multirow{2}{*}{2} & \BartCure(1) & 0.577 & 0.570 & 0.013 & 0.563\\
 & & CSF & 0.573 & 0.551 & 0.038 & 0.529\\
\cmidrule(l{3pt}){2-7}
 & \multirow{2}{*}{3} & \BartCure(1) & 0.197 & 0.188 & 0.046 & 0.179\\
 & & CSF & 0.091 & 0.079 & 0.127 & 0.068\\
\cmidrule(l{3pt}){2-7}
 & \multirow{2}{*}{4} & \BartCure(1) & 0.326 & 0.319 & 0.020 & 0.313\\
 & & CSF & 0.338 & 0.330 & 0.026 & 0.321\\
\bottomrule
\end{tabular}
}
\centering
\caption{Directional sign recovery for centered individual RMST heterogeneity.\label{tab:types}}
\end{table}

Table~\ref{tab:types} shows that \BartCure\ is generally more effective at recovering the direction of RMST heterogeneity. In the Henderson settings, \BartCure\ makes substantially more discoveries than CSF, with discovery rates around $35\%$ compared to around 6\%, while keeping Type-S error rates small. This leads to much larger net directional scores for \BartCure\ across all four Henderson DGPs. The same pattern holds for Cui~2 and Hu~3, where \BartCure\ has both a higher correct sign rate and lower Type-S error rates than CSF.

The main exceptions to this trend occur at Cui~1, Hu~1, and Hu~4. For Cui~1, both methods make very few discoveries, and the net directional scores are close to zero, indicating that neither method reliably detects directional heterogeneity; given the small amount of treatment effect heterogeneity in this setting, both methods appear to be behaving reasonably in failing to detect heterogeneity across the board. For Hu~1 and Hu~4, CSF is slightly more aggressive and obtains marginally larger net directional scores, although this comes with higher Type-S error rates. Overall, these results suggest that \BartCure\ is conservative in weak-signal settings but, when heterogeneity is identifiable, it tends to recover directional effects with a favorable balance between discovery and sign accuracy.

\section{Application to CALGB~40101 Trial}

We now apply \BartCure\ to data from 3,871 participants in the CALGB~40101 trial, a Phase~III randomized trial designed to evaluate the noninferiority of paclitaxel (T) relative to cyclophosphamide+doxorubicin (CA) for the treatment of breast cancer. Our primary endpoint is disease-free survival (DFS). Covariates of interest include age, race, ethnicity, menopause status (\texttt{stra1}), hormone receptor status (\texttt{stra2}), tumor size, assigned treatment duration, and number of positive lymph nodes. We restrict attention to 3,864 subjects with complete covariate information. As suggested by \citet[][Section 6.2.4]{peng2014cure}, we set $\tau$ in Assumption~5 to just-after the last observed failure time ($\tau = 115$ months).

\paragraph{Average Effects and Model Comparison}

\begin{figure}[t]
\centering
\includegraphics[width=0.75\textwidth]{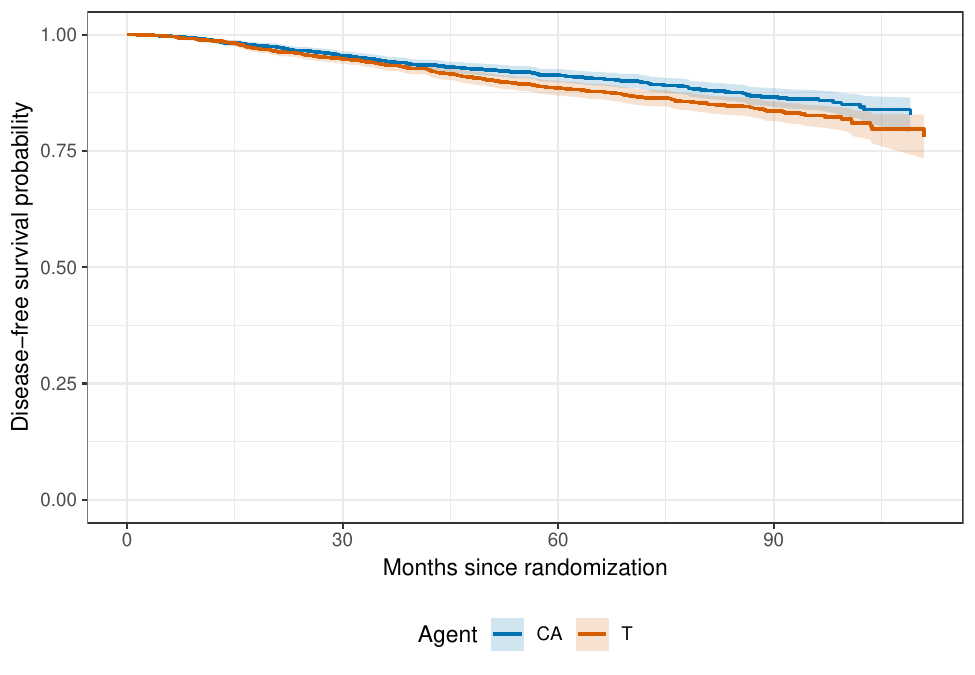}
\caption{Kaplan--Meier disease-free survival curves by treatment agent, with pointwise confidence bands.}
\label{fig:pds-agent-survival}
\end{figure}

Table~\ref{tab:rmst-compare} compares average causal effect estimates from different models for RMST and cure probability. Because treatment assignment is randomized, estimates based on the Kaplan--Meier estimator (displayed in Figure~\ref{fig:pds-agent-survival}) provide a useful benchmark. The Kaplan--Meier-based approach estimates a $-2.39$-month RMST difference (95\% CI: $-4.31$, $-0.47$) and a $-4.7\%$ cure-probability difference (95\% CI: $-10.7\%$, $1.3\%$). \BartCure\ is closest to this benchmark for $\Delta_R(115)$ ($-2.24$ months), with intervals excluding zero for both $\Delta_R(115)$ and $\Delta_C$; this is sensible because, while \BartCure\ is nonparametric, it applies some amount of shrinkage to the treatment effect. CSF estimates an RMST contrast in the same direction and of slightly larger magnitude, but with much wider intervals that include zero; its cure-probability estimate is also directionally similar but imprecise. IndivAFT gives the most attenuated estimates for both RMST ($-1.94$ months) and cure probability ($-2.8\%$), with comparatively narrow intervals; this may be due either to the imposition of the accelerated failure time assumption or to the fact that IndivAFT does not account for the existence of a cured population. Overall, the model-based methods agree on the direction of the effect, but \BartCure\ most closely reproduces the nonparametric RMST estimate.

\begin{table}
  \centering
  \scalebox{.97}{
  \begin{tabular}[t]{lrrrrrrrr}
    \toprule
                  & \multicolumn{4}{c}{$\widehat{\Delta}_R(115)$} & \multicolumn{4}{c}{$\widehat{\Delta}_C$}                             \\
    \cmidrule(lr){2-5} \cmidrule(lr){6-9}
    Method        & Estimate                                      & LCL    & UCL    & $P$-value & Estimate & LCL    & UCL    & $P$-value \\
    \midrule
    Nonparametric & -2.389                                        & -4.308 & -0.470 & 0.0147    & -0.047  & -0.107 & 0.013  & 0.127     \\
    CSF           & -3.393                                        & -7.715 & 0.930  & 0.1240    & -0.079   & -0.179 & 0.022  & 0.127     \\
    IndivAFT      & -1.940                                        & -3.863 & -0.063 & 0.0432    & -0.028   & -0.057 & -0.001 & 0.043     \\
    \BartCure     & -2.241                                        & -4.137 & -0.334 & 0.0264    & -0.039   & -0.073 & -0.004 & 0.026     \\
    \bottomrule
  \end{tabular}
  }
  \caption{Results for the causal effects $\widehat \Delta_R(115)$ and $\widehat
    \Delta_C$ for the CALGB 40101 trial, where LCL and UCL denote the lower and
    upper limits of 95\% confidence/credible intervals.}
  \label{tab:rmst-compare}
\end{table}

\begin{figure}[t]
  \centering
  \includegraphics[width=\textwidth]{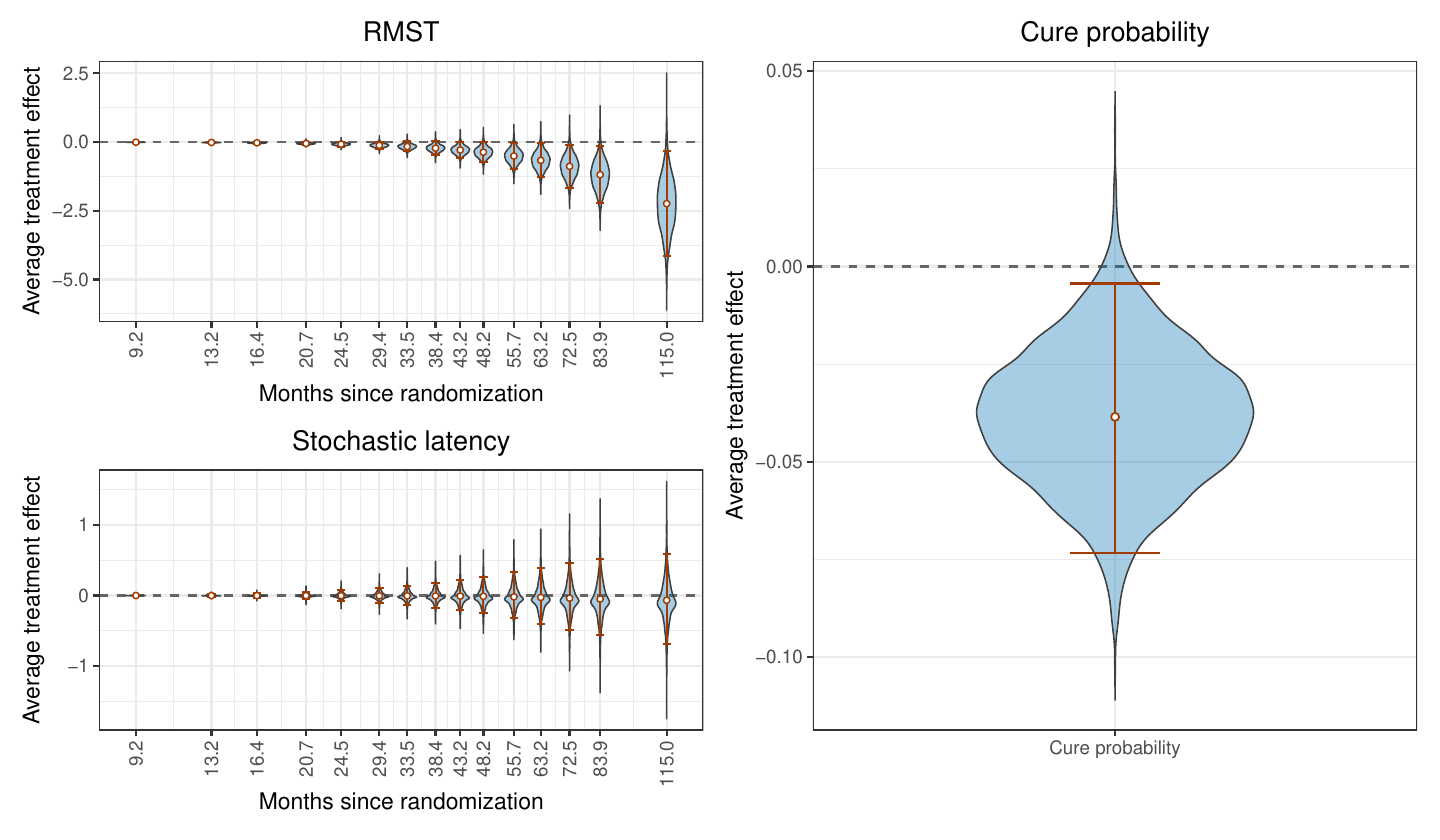}
  \caption{Posterior distributions of the RMST treatment effect, the stochastic latency effect, and the cure-probability treatment effect in the CALGB 40101 analysis over time. Points denote posterior means and vertical intervals denote 95\% credible intervals.}
  \label{fig:pds-stochastic-effects}
\end{figure}

The posterior distributions of $\Delta_R(t)$, $\Delta_{SL}(t)$, and $\Delta_C$ for \BartCure\ are given in Figure~\ref{fig:pds-stochastic-effects}. Because $\Delta_{SL}(t)$ is indistinguishable from $0$, the RMST effect appears primarily attributable to T curing fewer individuals rather than to latency differences among uncured individuals

\begin{figure}[t]
\centering
\includegraphics[width=0.95\textwidth]{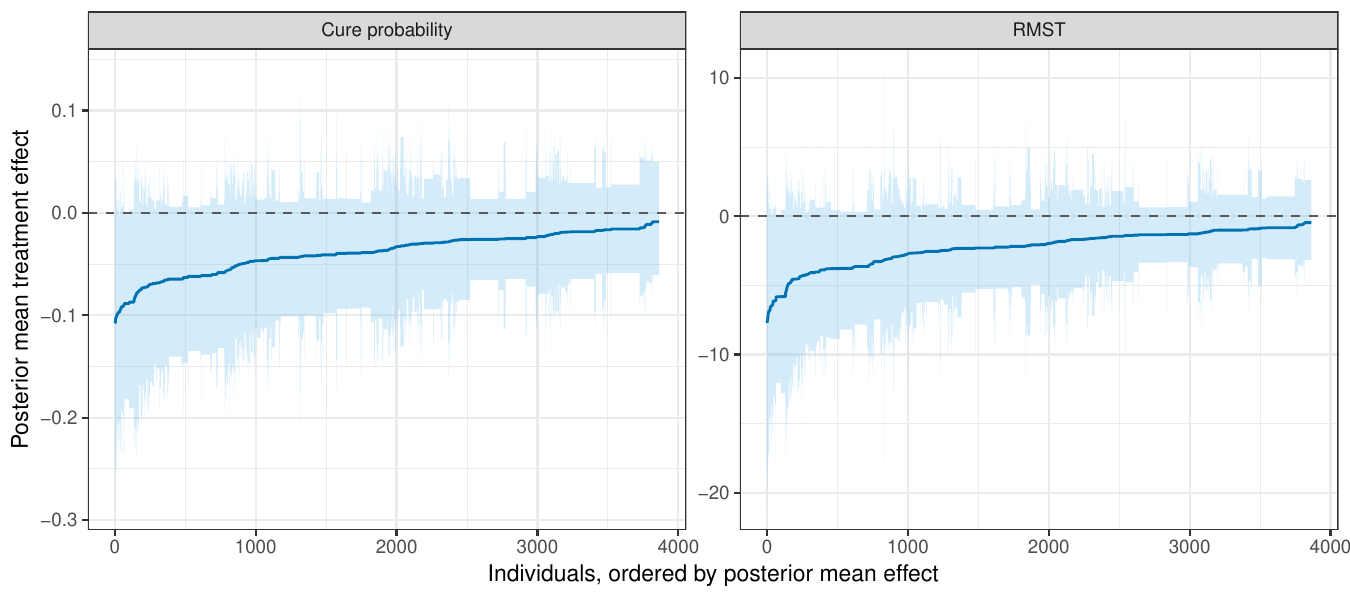}
\caption{Waterfall plots of individual posterior treatment effects for $\Delta_R(115, X_i)$ and $\Delta_C(X_i)$.}
\label{fig:pds-waterfalls}
\end{figure}

\paragraph{Individual Effects}
The posterior distributions of $\Delta_R(115, X_i)$ and $\Delta_C(X_i)$ for the individuals in the sample are given in Figure~\ref{fig:pds-waterfalls}; individuals are ordered by the posterior mean effect, and bands show pointwise 95\% credible intervals for the effects. We see that RMST and cure probability are very closely related. All individual effect estimates favor CA over T, with differing amounts of supporting evidence. Individual posterior probabilities of negative effects of T are given in Table~\ref{tab:pds-negative-effect-prob}; we see that roughly 75\% of individuals are estimated to have a posterior probability of a negative treatment effect of at least 80\%, and around 18\% have at least a 95\% posterior probability of a negative effect.

\begin{table}[t]
\centering
\begin{tabular}[t]{lll}
\toprule
Pr(negative effect) & Cure probability & RMST\\
\midrule
{}[0.50, 0.80) & 25.8\% & 24.0\%\\
{}[0.80, 0.90) & 22.3\% & 22.4\%\\
{}[0.90, 0.95) & 33.7\% & 35.8\%\\
{}[0.95, 0.99) & 18.1\% & 17.8\%\\
\bottomrule
\end{tabular}
\caption{Distribution of individual posterior probabilities of a negative treatment effect.}
\label{tab:pds-negative-effect-prob}

\end{table}

\begin{figure}[t]
\centering
\includegraphics[width=0.65\textwidth]{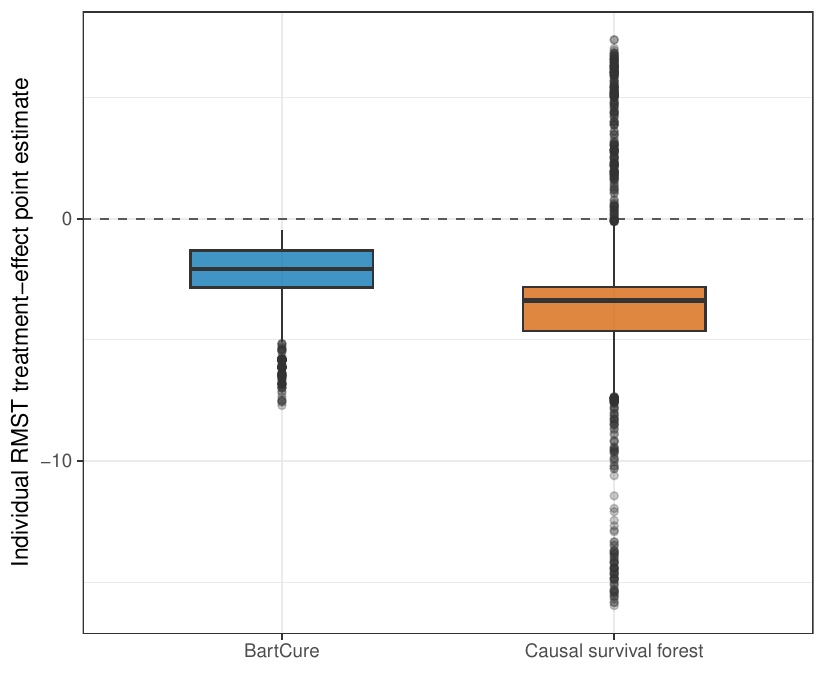}
\caption{Boxplots of individual RMST treatment-effect point estimates from \BartCure\ and the causal survival forest.}
\label{fig:pds-rmst-effect-boxplot}
\end{figure}

For comparison, boxplots of individual RMST effect estimates for \BartCure\ and CSF are given in Figure~\ref{fig:pds-rmst-effect-boxplot}. We see that CSFs produce much larger estimates of treatment effect heterogeneity, with many individuals estimated to have a positive T-versus-CA effect. By contrast, \BartCure\ regularizes towards a nearly-homogeneous treatment effect.

\begin{figure}[t]
  \centering
  \includegraphics[width=.45\textwidth]{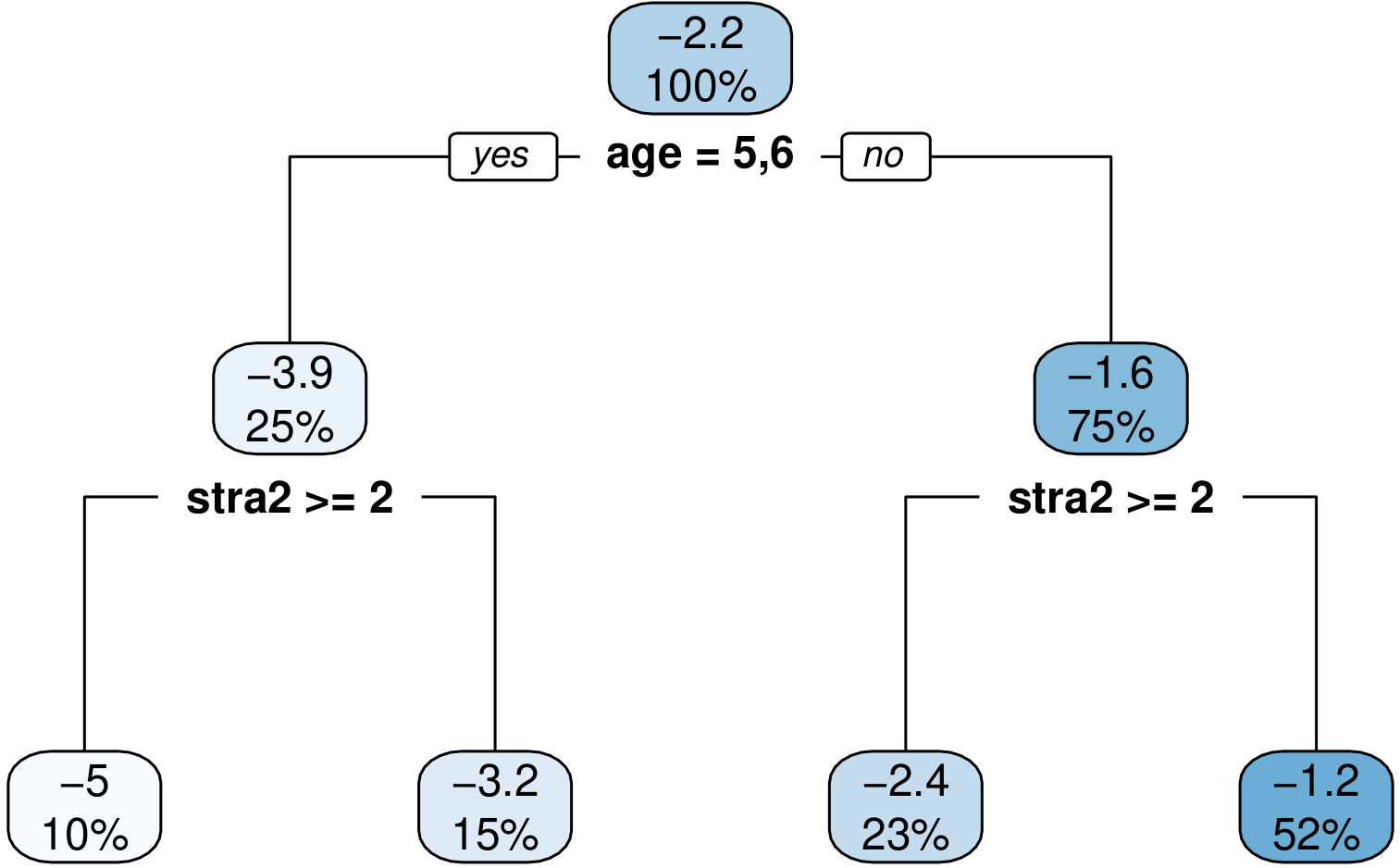}
  \caption{Tree-based summary of posterior heterogeneity in the \BartCure\ RMST treatment effect. Numbers in the nodes give estimated subgroup RMST effect and the subgroup size.}
  \label{fig:bartcure_tree}
\end{figure}

\paragraph{Subgroup Effects}
Following \citet{hahn2020bayesian,alam2025decision}, we constructed a decision tree summary to find optimal decision-theoretic subgroups that maximize treatment effect heterogeneity. Figure~\ref{fig:bartcure_tree} provides a decision tree summary of $\Delta_R(t,x)$. This summary identifies two variables that the treatment may differ across: the age of an individual (which was stratified into 6 groups, with groups 5 and 6 corresponding to individuals above the age of 60), and \texttt{stra2} (which stratifies on hormone receptor status, with 1 for positive and 2 for negative). Generally, older individuals with negative receptor status have more negative RMST contrast estimates across arms. 

\begin{figure}
  \centering
  \includegraphics[width=1\textwidth]{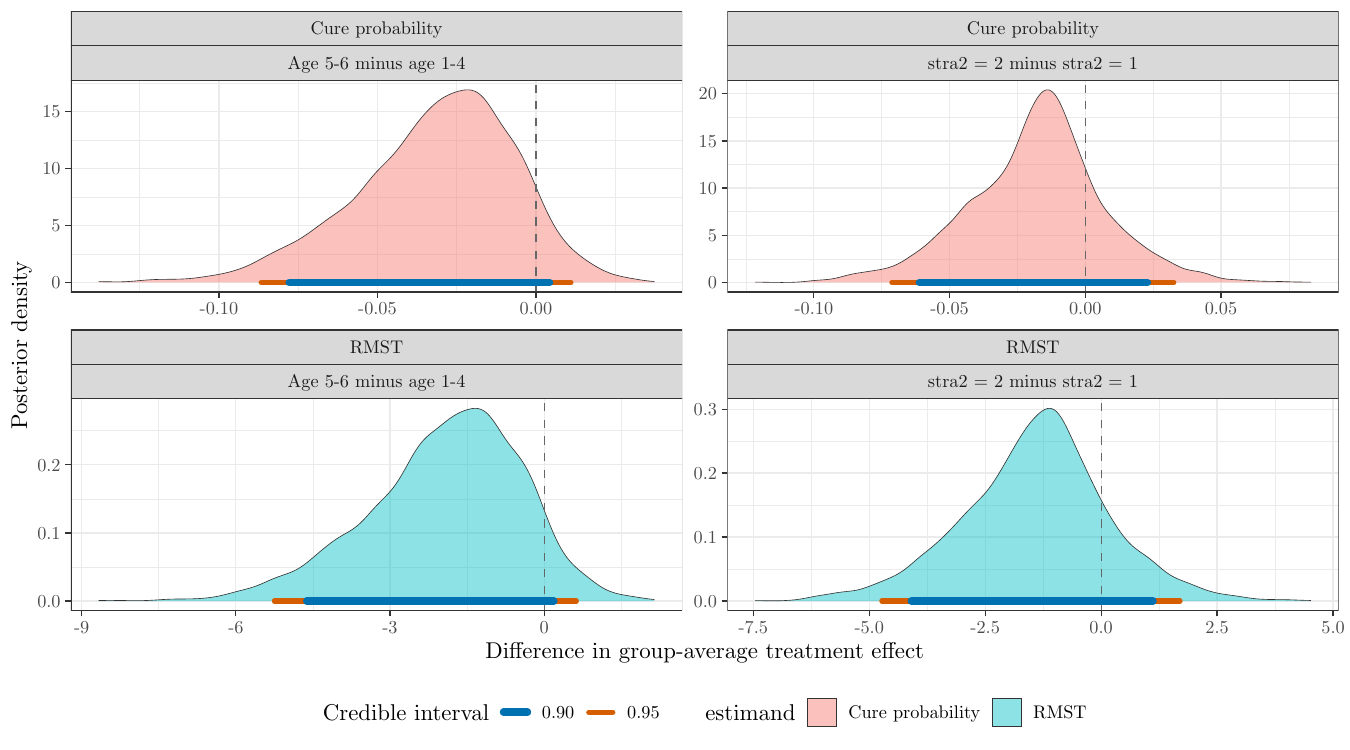}
  \includegraphics[width=1\textwidth]{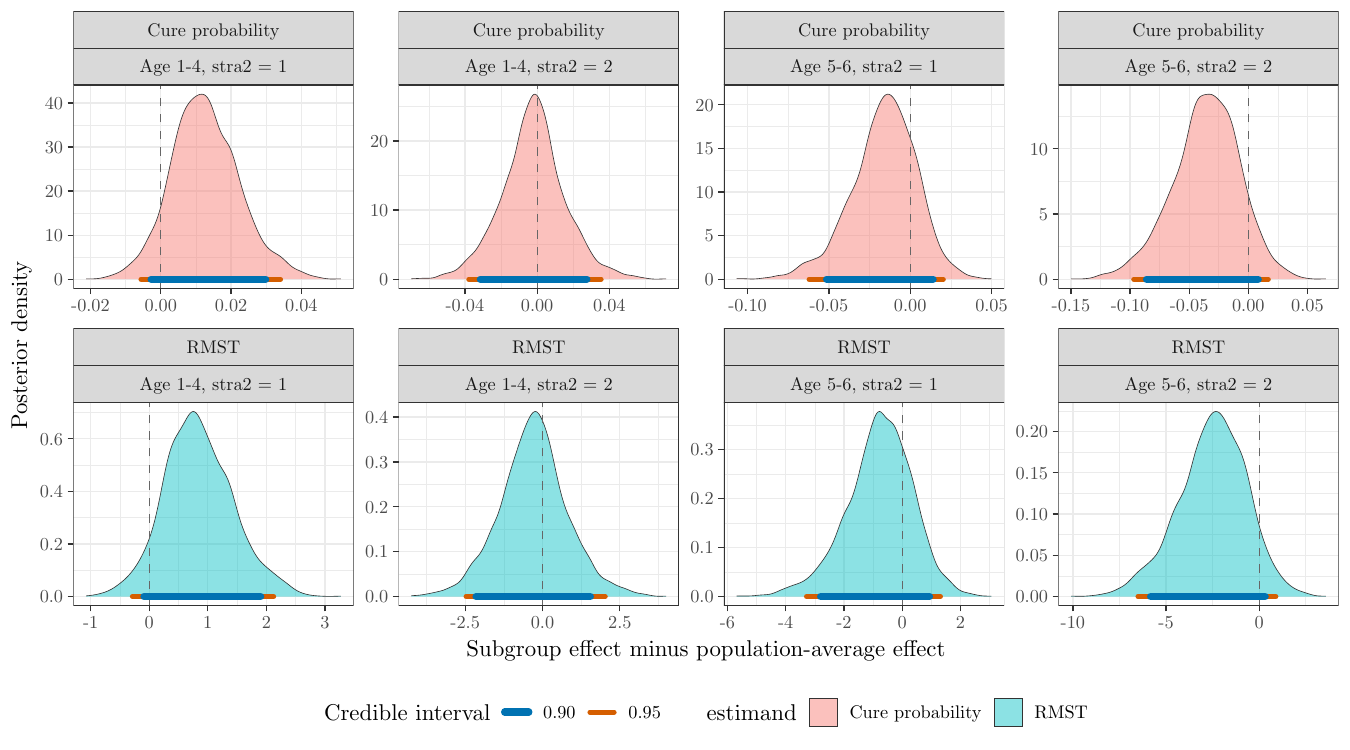}
  \caption{Posterior densities for differences in group-average treatment effects at 115 months, comparing age groups and \texttt{stra2} (1 if receptor status is positive and 2 if negative). Horizontal bars denote 90\% and 95\% posterior credible intervals.}
  \label{fig:analyze_project_data_sphere_groups}
\end{figure}

Figure~\ref{fig:analyze_project_data_sphere_groups} shows posterior distributions for $\Delta_R(t)$ and $\Delta_C$ within these subgroups, centered by the population average treatment effect. For all subgroups, there is at-most weak evidence supporting treatment effect heterogeneity, with sizeable uncertainty. We see that older individuals have an RMST difference estimated to be roughly 1.5 months lower and a cure probability effect about 2.5\% lower than younger individuals. We also see a negative effect of \texttt{stra2 = 2}, and both of these effects are roughly additive when looked at jointly.

\paragraph{Additional Results}
Further analysis of this dataset is available in the Supplementary Material. This includes an assessment of variable importance and analysis of the proportion of the RMST effect attributable to cure.

\section{Discussion}

In this work, we proposed a causal machine learning framework for survival settings with a cured subpopulation. We defined causal effects on finite-time survival, restricted mean survival time (RMST), and cure probability, and considered decompositions of the RMST effect into contributions arising from cure and from delayed failure among uncured individuals. This cure-rate formulation provides clinically relevant information that standard causal analyses do not directly capture. We first considered a naive latency contrast, which is a well-defined causal estimand, but argued that its interpretation is unsatisfactory. To address this issue, we introduced stochastic cure and latency effects that separate the contribution of the cure probability from that of the finite-event-time distribution. We also showed how the stochastic latency effect relates to a principal-strata effect among individuals who would remain uncured under either treatment arm. For estimation, we proposed \BartCure, a BART-based promotion-time cure model targeting both average and conditional versions of these effects.

The main strength of \BartCure\ is its ability to estimate clinically interpretable cure-rate causal effects while retaining the flexibility of Bayesian nonparametric regression. In simulations, \BartCure\ performs well relative to causal survival forests and an accelerated failure time BART model, both in settings with a cured subpopulation and in settings where all survival times are finite. A second strength is its behavior when estimating heterogeneous treatment effects. In many applications, overstating heterogeneity can be more harmful than failing to detect weak heterogeneity, since subgroup claims often influence clinical interpretation and the design of future studies. The prior used by \BartCure\ shrinks treatment effects toward homogeneity, and the simulation results suggest that this induces conservative heterogeneity detection with favorable Type-S error properties. Consequently, when the method identifies strong subgroup patterns, those patterns have persisted despite the model's regularization toward homogeneous effects.

The proposed estimands should be interpreted with care. The stochastic latency effect is identified as a functional of the observed data distribution under the stated assumptions, but connecting it to the principal-strata effect $\Delta_{UU}$ requires additional conditions; the connection is most direct under monotonicity and a principal ignorability condition relating the control-arm RMST distribution in the $UU$ and $UC$ strata. These assumptions may be plausible in some clinical settings, but they are not fully testable. A second limitation is the reliance on a cure-threshold assumption. This assumption is common for semiparametric cure models, and is useful for stabilizing estimation of the cured fraction, but it remains a substantive modeling choice. If the threshold is chosen too small, we may incorrectly classify uncured individuals as cured; if it is chosen too large, the cure probability may be weakly identified in small samples. In practice, the threshold should therefore be guided by clinical knowledge and sensitivity analyses should be conducted whenever possible.

\paragraph{Acknowledgements}
This work was supported under NSF grant DMS-2144933.

\paragraph{Supplementary Material}
Supplementary files include formal identification assumptions, proofs of all propositions, full details of the Gibbs sampling algorithm used to fit the models, additional simulation details and experiments, additional analyses CALGB 40101, and code reproducing figures and tables.

\bibliographystyle{apalike}
\bibliography{mybib,references}

\clearpage

\appendix
\clearpage
\pagestyle{fancy}
\fancyhf{}
\fancyhead[L]{\textsc{Supplementary Material}}
\fancyhead[R]{\thepage}
\renewcommand{\headrulewidth}{0.4pt}

\renewcommand{\thesection}{S\arabic{section}}
\renewcommand{\thesubsection}{S\arabic{section}.\arabic{subsection}}
\renewcommand{\thefigure}{S\arabic{figure}}
\renewcommand{\thetable}{S\arabic{table}}
\renewcommand{\theequation}{S\arabic{equation}}
\renewcommand{\theproposition}{S\arabic{proposition}}
\renewcommand{\theHsection}{supp.\arabic{section}}
\renewcommand{\theHsubsection}{supp.\arabic{section}.\arabic{subsection}}
\renewcommand{\theHfigure}{supp.\arabic{figure}}
\renewcommand{\theHtable}{supp.\arabic{table}}
\renewcommand{\theHequation}{supp.\arabic{equation}}
\renewcommand{\theHproposition}{supp.\arabic{proposition}}
\setcounter{section}{0}
\setcounter{subsection}{0}
\setcounter{figure}{0}
\setcounter{table}{0}
\setcounter{equation}{0}
\setcounter{proposition}{0}

\begin{center}
  {\Large\bfseries Supplementary Material\par}
  \vspace{0.5em}
  {\large Bayesian Causal Machine Learning for Cure Models\par}
\end{center}

\section{Causal Identification Assumptions}
\label{sec:supp-causal-identification}

We require the following assumptions to identify the causal parameters. Assumptions 1--3 are standard assumptions in the causal inference literature \citep{rubin2005causal}, which state that the treatment is well-defined with no interference between units, all individuals have some non-negligible probability of being assigned to either treatment or control, and there are no unmeasured common causes of the treatment assignment mechanism and the potential outcomes; in CALGB 40101, the treatment assignment was randomized, and so these assumptions are known to hold. Assumption 4 is also standard in survival analysis, and is plausible when censoring is primarily administrative.

\begin{description}
\item[Assumption 1:] \textbf{Stable Unit Treatment Value (SUTVA).}
  The observed failure time satisfies $T_i = T_i(A_i)$, and the potential
  outcomes of individual $i$ do not depend on the treatment assignments of
  other individuals \citep{rubin1980randomization}.

\item[Assumption 2:] \textbf{Positivity.}
  There exists a $\delta > 0$ such that
  $\delta < e(x) < 1 - \delta$ for all $x$ in the
  support of $X_i$, where we recall that $e(x) = \Pr(A_i = 1 \mid X_i = x)$ is the propensity score.

\item[Assumption 3:] \textbf{Ignorability.}
  Treatment assignment is independent of the potential outcomes given the
  observed covariates:
  \begin{math}
    \{T_i(0), T_i(1)\} \perp A_i \mid X_i.
  \end{math}

\item[Assumption 4:] \textbf{Ignorable Censoring.}
  The censoring mechanism is independent of the observed failure times given
  the treatment and covariates:
  \begin{math}
    C_i \perp T_i \mid A_i, X_i.
  \end{math}

\item[Assumption 5:] \textbf{Cure Threshold.} There exists a known time $\tau < \infty$ such that the hazard function satisfies $h(t \mid a, x) = 0$ for all $t > \tau$, $a \in \{0,1\}$, and $x$ in the support of $X_i$. Equivalently, survival beyond $\tau$ implies cure. Additionally, we require sufficient follow-up in the sense that $\Pr(C_i > \tau \mid X_i = x) > 0$ for all $x$.
\end{description}

\section{Proofs of Propositions}
\label{sec:supp-proofs}

\paragraph{Proof of Proposition~\ref{prop:uu}.}
Because $L_i(a,t) = R_i(a,t)1\{T_i(a) < \infty\}$,
\begin{align*}
  \Delta_L(t)
  &=
  \E\left[R_i(1,t)1\{T_i(1) < \infty\}
  -
  R_i(0,t)1\{T_i(0) < \infty\}\right].
\end{align*}
The condition $\Pr(i \in CU \mid X_i) = 0$ implies $\Pr(i \in CU)=0$, so $1\{T_i(1)<\infty\}=1\{i \in UU\}$ and $1\{T_i(0)<\infty\}= 1\{i \in UU \cup UC\}$. Hence
\begin{align*}
  \Delta_L(t)
  &=
  \Pr(i \in UU)\E\{R_i(1,t)-R_i(0,t)\mid i \in UU\}
  -
  \Pr(i \in UC)\E\{R_i(0,t)\mid i \in UC\}.
\end{align*}
Under the same condition, $p_1=\Pr(i\in UU)$ and
$\Delta_C=\Pr\{T_i(1)=\infty\}-\Pr\{T_i(0)=\infty\}
=\Pr(i\in UC)=p_0-p_1$. Substitution gives the result.

\paragraph{Proof of Proposition~\ref{prop:stochastic-decomposition}.}
Fix $x$. By definition,
\begin{align*}
  \Delta_R(t,x)
  &=
  \{1-p_1(x)\}t + p_1(x)m_1(t,x)
  -
  \{1-p_0(x)\}t - p_0(x)m_0(t,x) \\
  &=
  \{p_0(x)-p_1(x)\}t
  +
  p_1(x)m_1(t,x)-p_0(x)m_0(t,x).
\end{align*}
Also,
\begin{align*}
  \Delta_{SC}(t,x)+\Delta_{SL}(t,x)
  &=
  \{p_0(x)-p_1(x)\}\left\{t-\frac{m_0(t,x)+m_1(t,x)}{2}\right\} \\
  &\quad+
  \{m_1(t,x)-m_0(t,x)\}\left\{\frac{p_0(x)+p_1(x)}{2}\right\} \\
  &=
  \{p_0(x)-p_1(x)\}t
  +
  p_1(x)m_1(t,x)-p_0(x)m_0(t,x),
\end{align*}
which equals $\Delta_R(t,x)$.

\paragraph{Proof of Proposition~\ref{prop:uu2}.}
First, note that because $\Pr(i \in CU \mid X_i = x) = 0$, an individual who is uncured under treatment must belong to the stratum \(UU\). Hence
\begin{math}
  \Pr\{T_i(1)<\infty \mid X_i=x\}
  =
  \Pr(i\in UU\mid X_i=x),
\end{math}
so that $\Pr(i\in UU\mid X_i=x)=p_1(x)$. Similarly, an individual who is uncured under control belongs to either $UU$ or $UC$, and therefore
\begin{math}
  p_0(x) = \Pr(i\in UU\mid X_i=x) + \Pr(i\in UC\mid X_i=x).
\end{math}
It follows that $\Pr(i\in UC\mid X_i=x) = p_0(x)-p_1(x) = \Delta_C(x)$.

Now consider the conidtional RMST among uncured individuals. Because $T_i(1) < \infty$ is equivalent to $i \in UU$, we have
\begin{align*}
  m_1(t,x)
  =
  \E\{R_i(1,t)\mid i\in UU, X_i=x\}.
\end{align*}
On the other hand, rearranging the expression $\E\{R_i(a,t) \mid X_i = x\} = \{ 1 - p_a(x)\} \times t + p_a(x) \times m_a(t,x)$ with $a = 0$ and applying $\E\{R_i(0,t) \mid X_i = x\} = \Pr(i \in UU \mid X_i = x) \times \E\{R_i(0,t) \mid i \in UU, X_i = x\} + \Pr(i \in UC \mid X_i = x) \times \E\{R_i(0,t) \mid i \in UC, X_i = x\}$ gives
\begin{align*}
  m_0(t,x) &=
  \frac{
    p_1(x)\E\{R_i(0,t)\mid i\in UU, X_i=x\}
    +
    \Delta_C(x)\E\{R_i(0,t)\mid i\in UC, X_i=x\}
  }{p_0(x)}.
\end{align*}
Using the notation
\begin{math}
  B_{UU}(t,x)=\E\{R_i(0,t)\mid i\in UU, X_i=x\},
  B_{UC}(t,x)=\E\{R_i(0,t)\mid i\in UC, X_i=x\},
\end{math}
and
\begin{math}
  \Delta_{UU}(t,x)
  =
  \E\{R_i(1,t)-R_i(0,t)\mid i\in UU, X_i=x\},
\end{math}
we can write
\begin{align*}
  m_1(t,x)=B_{UU}(t,x)+\Delta_{UU}(t,x)
  \quad \text{and} \quad
  m_0(t,x)
  =
  \frac{p_1(x)B_{UU}(t,x)+\Delta_C(x)B_{UC}(t,x)}{p_0(x)}.
\end{align*}
Therefore,
\begin{align*}
  m_1(t,x)-m_0(t,x)
  &=
  B_{UU}(t,x)+\Delta_{UU}(t,x)
  -
  \frac{p_1(x)B_{UU}(t,x)+\Delta_C(x)B_{UC}(t,x)}{p_0(x)} \\
  &=
  \Delta_{UU}(t,x)
  +
  \frac{\{p_0(x)-p_1(x)\}B_{UU}(t,x)-\Delta_C(x)B_{UC}(t,x)}{p_0(x)} \\
  &=
  \Delta_{UU}(t,x)
  +
  \frac{\Delta_C(x)}{p_0(x)}
  \{B_{UU}(t,x)-B_{UC}(t,x)\}.
\end{align*}
Finally, by definition,
\[
  \Delta_{SL}(t,x)
  =
  \{m_1(t,x)-m_0(t,x)\}
  \frac{p_0(x)+p_1(x)}{2}.
\]
Substituting the preceding display gives the stated identity. If either \(\Delta_C(x)=0\) or \(B_{UU}(t,x)=B_{UC}(t,x)\), the second term in braces vanishes, so \(\Delta_{SL}(t,x)\) is proportional to \(\Delta_{UU}(t,x)\). This proves the final claim.


\section{Full Details of Gibbs Sampler}
\label{sec:supp-gibbs}

For the one-forest model, the Gibbs sampler used is exactly the one described by \citet{alam2025unified}, with the additional constraint that $\lambda_K \equiv 0$ is enforced at the end of each update of the $\lambda$'s.

For the two-forest model, consider the promotion-time survival model with survival function
\begin{align*}
  S(t \mid x) = \exp\left\{ -\theta(x) \Ft(t \mid x) \right\}.
\end{align*}
This model has hazard function
\begin{align*}
  h(t \mid x) = \theta(x) \, \ft(t \mid x).
\end{align*}
Adopt the shorthand $S = S(t \mid x)$, $\Ft = \Ft(t \mid x)$, $\theta = \theta(x)$, and $\ft = \ft(t \mid x)$; adding the subscript $i$ is shorthand for plugging in $t = Y_i$ and $x = X_i$ so $S_i = S(Y_i \mid X_i)$. We also write $\St = 1 - \Ft$ and $\htt = -\frac{\partial}{\partial t} \log \St$.

The likelihood in the promotion-time model is
\begin{align*}
  \prod_i (\theta_i \ft_i)^{\delta_i} \exp\left\{ -\int_0^{Y_i} \theta_i \, \ft \ dt \right\}
  =
  \prod_i \left( \theta_i \, \htt_i \, \St_i \right)^{\delta_i}
  \exp\left( -\theta_i \, \Ft_i \right).
\end{align*}
Now, augment $K_i \sim \Poisson(\theta_i \, \St_i)$. The augmented likelihood becomes
\begin{align*}
  \prod_i  \frac{\theta_i^{K_i + \delta_i} e^{-\theta_i}}{K_i!}
  \times \htt_i^{\delta_i} \St_i^{\delta_i + K_i}.
\end{align*}
Next, we consider the model 
\begin{math}
  \log \theta = r^{\theta}(x),
\end{math}
with shorthand $r^\theta$ and $r^\theta_i$. We also consider the model $\htt = \lambda_{B(t)} \exp\{r^h(B(t), x)\}$, also with shorthand $\htt = \lambda \, \exp(r^h)$. Both of these terms can be handled in a fully conjugate fashion using log-linear BART models with log-gamma priors. For example, we could set
\begin{align*}
  r^\theta = \sum_{m = 1}^M g(x; \Tree^\theta_m, \sM^\theta_m)
\end{align*}
with leaf node parameters $\mu^\theta_{\ell m} \sim \log \Gam(a^\theta, b^\theta)$. Similarly, we can use the model
\begin{math}
  r^h(B(t), x) = r^h = \sum_{m = 1}^M g(B(t), x ; \Tree^r_m, \sM^r_m)
\end{math}
with leaf node parameters $\mu^r_{\ell m}$; the interpretation here is that we have $K_i$ censored values above $Y_i$ in the usual piecewise exponential model, with $Y_i$ itself either denoting an event time $(\delta_i = 1)$ so that we have observed an event exactly at $Y_i$, or itself might correspond to censoring (so that we have $K_i + 1$ events to consider above $Y_i$; this shouldn't be interpreted literally in the context of the cure model, but is just how one can interpret this component of the likelihood).

The Poisson log-linear model updates are described by \citet{murray2021log} and, while the survival updates are also relatively straightforward given the above development, we will also outline them here.

Fix leaf $\ell$ of tree $m$ and define $\eta_{ib} = r^h(b, X_i) - \mu_{\ell m}$. Define $Z_{ib} = 1(Y_i \ge c_b) (c_b - c_{b-1}) + 1(c_{b-1} \le Y_i < c_b)(Y_i - c_{b-1})$ and $\delta_{ib} = 1(\delta_i = 1 \wedge Y_i \in (c_{b-1}, c_b))$. Then the likelihood associated to this leaf node is 
\begin{align*}
  \prod_{i, b : (b, X_i) \leadsto (\ell, m)}
  \exp\left\{ \delta_{ib} (\eta_{ib} + \mu_{\ell m})
  - (\delta_i + K_i)\lambda_b \, Z_{ib} \, e^{\eta_{ib}} e^{\mu_{\ell m}}
  \right\},
\end{align*}
where $(b,X_i) \leadsto (\ell, m)$ denotes that $(b,X_i)$ is associated to leaf node $\ell$ of tree $m$. Integrating out $\mu_{\ell m}$ against its prior gives the integrated likelihood
\begin{align*}
  m(\Tree_m) = \prod_{\ell}
  \frac{\Gamma(a_\mu + A_\ell)}{(b_\lambda + B_\ell)^{a_\mu + A_\ell}} \times \frac{b_\mu ^{a_\mu }}{\Gamma(a_\mu)}
\end{align*}
where $A_\ell = \sum_{(i,b) : (b, X_i) \leadsto (\ell, m)} \delta_{ib}$ and $B_\ell = \sum_{(i,b) : (b, X_i) \leadsto (\ell, m)} (\delta_i + K_i) \lambda_b Z_{ib} e^{\eta_{ib}}$. This integrated likelihood can be used, as described by \citet{hill2020bayesian}, as the basis for a Metropolis-Hastings update for $\Tree_\ell$. Given a proposal distribution $\Lambda(\Tree \mid \Tree')$, we accept a proposed $\Tree' \sim \Lambda(\Tree \mid \Tree_m)$ with probability
\begin{align*}
  1 \wedge \frac{m(\Tree') \, \pi_\Tree(\Tree') \, \Lambda(\Tree_m \mid \Tree')}{m(\Tree_m) \, \pi_\Tree(\Tree_m) \, \Lambda(\Tree' \mid \Tree_m)}.
\end{align*}
Reasonable proposals for new trees are the grow, prune, and change moves described by \citet{chipman1998bayesian}. After updating $\Tree_\ell$ by Metropolis-Hastings, we then sample $\mu_{\ell m} \sim \log \Gam(a_\mu + A_\ell, b_\mu + B_\ell)$.

Finally, after all of the $(\Tree_m, \sM_m)$'s have been updated, the baseline hazard parameters are updated similarly as
\begin{math}
  \lambda_b \sim \Gam(a_\lambda + A_b, b_\lambda + B_b)
\end{math}
where $A_b = \sum_i \delta_{ib}$ and $B_b = \sum_i (\delta_i + K_i) Z_{ib} e^{r^h_i}$.




\section{Additional Simulation Details}
\label{sec:supp-simulation-details}

\subsection{More Details on the DGPs}
\label{sec:supp-dgps}

We now provide the full details for the different DGPs used in the simulation experiments of Section~\ref{sec:experiments}.

\paragraph{\citet{henderson2020individualized}}
This setting is based on real data taken from the SOLVD trial \citep{henderson2020individualized}. Covariates and treatment data are taken directly from the trial data, while the outcome $T_i(a)$ follows an accelerated failure time model with $\log T_i(a) = \mu_a(x) + \epsilon_i$, where $\epsilon_i$ has mean zero and unit variance following either a normal (DGP~1), Gumbel (DGP~2), standardized gamma (DGP~3), or $t$-mixture (DGP~4) distribution. The propensity score is estimated by a logistic regression on $X_i$, while the censoring time is sampled independently as $C_i \sim \Uniform(6.5, 10)$. We use $t = 6$ for the follow-up time.

\paragraph{\citet{cui2023estimating}}
Covariates for this setting are sampled by setting $X_i = U_i V$ where $U_{ij} \sim \Uniform(0,1)$ and $V$ is given by the Cholesky decomposition of a matrix with $(j,k)^{\text{th}}$ entry $0.5^{|j - k|}$. We consider two sub-settings:
\begin{enumerate}
  \item Log-normal AFT: $\log T_i(a) = \mu_a(X_i) + \epsilon_i$ where $\epsilon_i \sim \Normal(0,1)$ and with $\mu_0(x) = -1.85 - 0.8 \cdot 1(x_1 < 0.5) + 0.7\sqrt{x_2} + 0.2 x_3$ and $\mu_1(x) = \mu_0(x) + 0.7 - 0.4 \cdot 1(x_1 < 0.5) - 0.4\sqrt{x_2}$. The propensity score is taken to be $e(x) = (1 + f_\beta(x_1; 2, 4))/4$, where $f_\beta$ is the Beta density. Censoring is covariate-dependent via a Weibull model with $C_i(a) = \sqrt{-\log(U_i)/\exp\{f_{Ca}(X_i)\}}$, where $U_i \sim \Uniform(0,1)$, $f_{C0}(x) = -1.75 - 0.5\sqrt{\max(x_2,0)} + 0.2x_3$, and $f_{C1}(x) = f_{C0}(x) + 1.15 + 0.5 \cdot 1(x_1 < 0.5) - 0.3\sqrt{\max(x_2,0)}$. The follow-up time is taken to be $t = 1.5$.
  \item Weibull proportional hazards:  $S(t \mid x, a) = \exp(-e^{f_a(x)} \, t^{1/2})$, with $f_0(x) = x_1$ and $f_1(x) = x_1 + x_2 - 0.5$. The propensity score is $e(x) = (1 + f_\beta(x_2; 2, 4))/4$, while censoring is $C_i \sim \Uniform(0, 3)$. The time horizon is $t = 1.25$.
\end{enumerate}

\paragraph{\citet{hu2021estimating}}
Covariates for this setting consist of five continuous variables $X_1,\ldots,X_5 \iid \Normal(0,0.35^2)$ and five binary variables $X_6,\ldots,X_{10} \iid \Bernoulli(0.5)$. The propensity score is $e(x) = \operatorname{expit}(0.3 - 0.25x_1 - 2.25x_2 - 0.75x_3 - 0.25x_5 - 0.25x_6 - 0.50x_7 - x_9 + 1.25x_{10})$. Potential event times follow Weibull proportional hazards models with shape $\eta = 2$ and scale parameters $\kappa_a(x) = d_a \exp\{f_a(x)\}$, where $d_0 = 1200$ and $d_1 = 2000$, so that $T_i(0) = \left[-\log(U_i) / \{1200\exp(f_0(X_i))\}\right]^{1/2}$ and $T_i(1) = \left[-\log(U_i) / \{2000\exp(f_1(X_i))\}\right]^{1/2}$ with $U_i \sim \Uniform(0,1)$. Censoring is independent with $C_i \sim \Gam(1, 0.007)$. The functions $f_0$ and $f_1$ are given by
\begin{enumerate}
\item DGP 1: $f_0(x) = 0.2 - 0.5x_1 - 0.8x_3 - 1.8x_5 - 0.9x_6 - 0.1x_7$ and $f_1(x) = -0.2 + 0.1\operatorname{expit}(x_1) - 0.8\sin(x_3) - 0.1x_5^2 - 0.3x_6 - 0.2x_7$.
\item DGP 2: $f_0(x) = -0.1 + 0.1x_1^2 - 0.2\sin(x_3) + 0.2\operatorname{expit}(x_5) + 0.2x_6 - 0.3x_7$ and $f_1(x) = -0.2 + 0.1\operatorname{expit}(x_1) - 0.8\sin(x_3) - 0.1x_5^2 - 0.3x_6 - 0.2x_7$.
\item DGP 3: $f_0(x) = -0.1 + 0.1x_1^2 - 0.2\sin(x_3) + 0.2\operatorname{expit}(x_5) + 0.2x_6 - 0.3x_7$ and $f_1(x) = 0.5 - 0.1\operatorname{expit}(x_2) + 0.1\sin(x_3) - 0.1x_4^2 + 0.2x_4 - 0.1x_5^2 + 0.2\operatorname{expit}(x_5) + 0.2x_6 - 0.3x_7$.
\item DGP 4: $f_0(x) = -0.2 + 0.5\sin(\pi x_1 x_3) + 0.2\operatorname{expit}(x_5) + 0.2x_6 - 0.3x_7$ and $f_1(x) = 0.5 - 0.1\operatorname{expit}(x_2) + 0.1\sin(x_3) - 0.1x_4^2 + 0.2x_4 - 0.1x_5^2 - 0.3x_6$.
\end{enumerate}
We use a time horizon of $t = 0.05$ for all settings.

\subsection{Results for Conditional Average Treatment Effects}
\label{sec:supp-cate-results}

\begin{figure}[p]
  \centering
  \includegraphics[width=\textwidth]{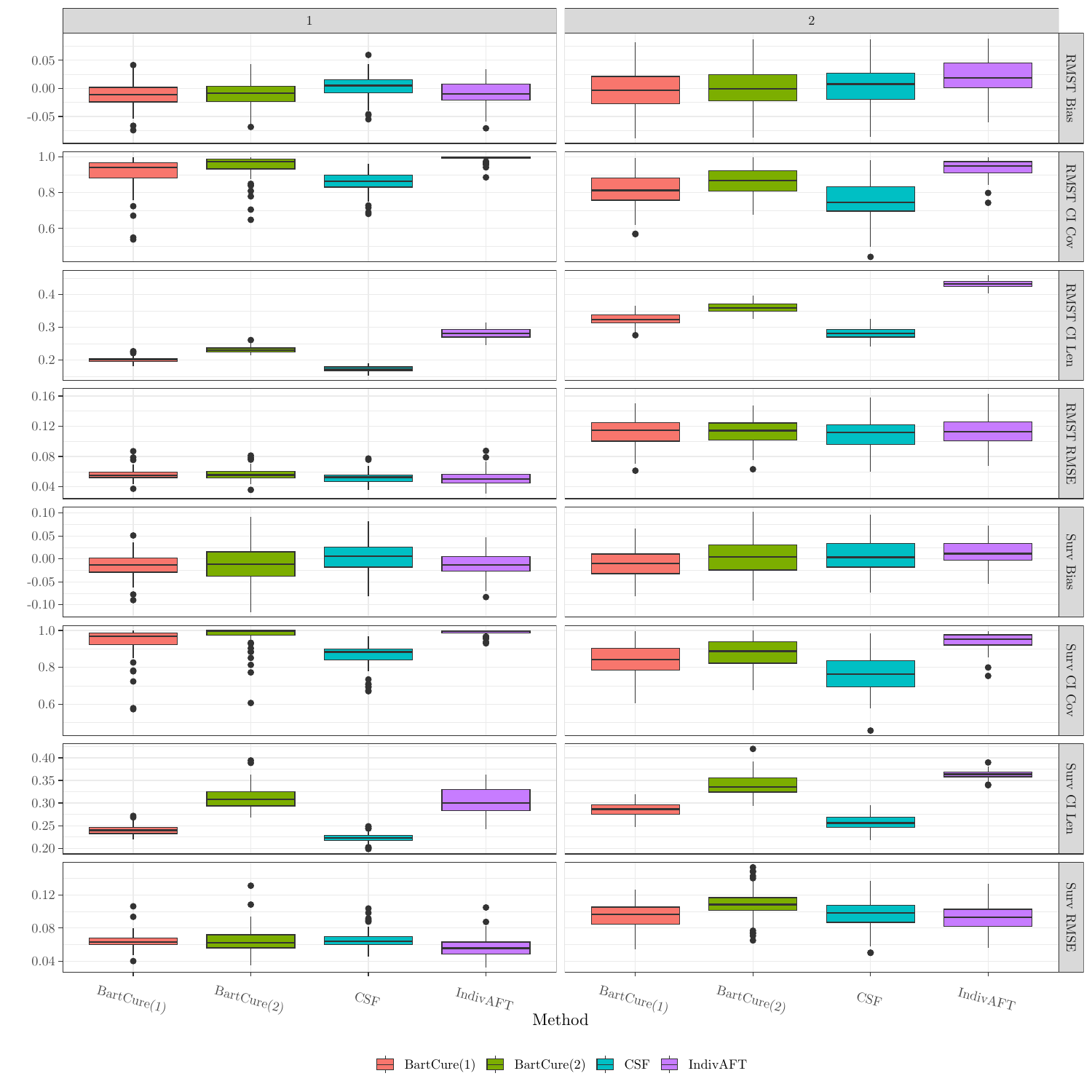}
  \caption{Conditional average treatment effect metrics for the Cui DGPs \emph{with} a cured fraction.\label{fig:cure_cui}}
\end{figure}

\begin{figure}[p]
  \centering
  \includegraphics[width=\textwidth]{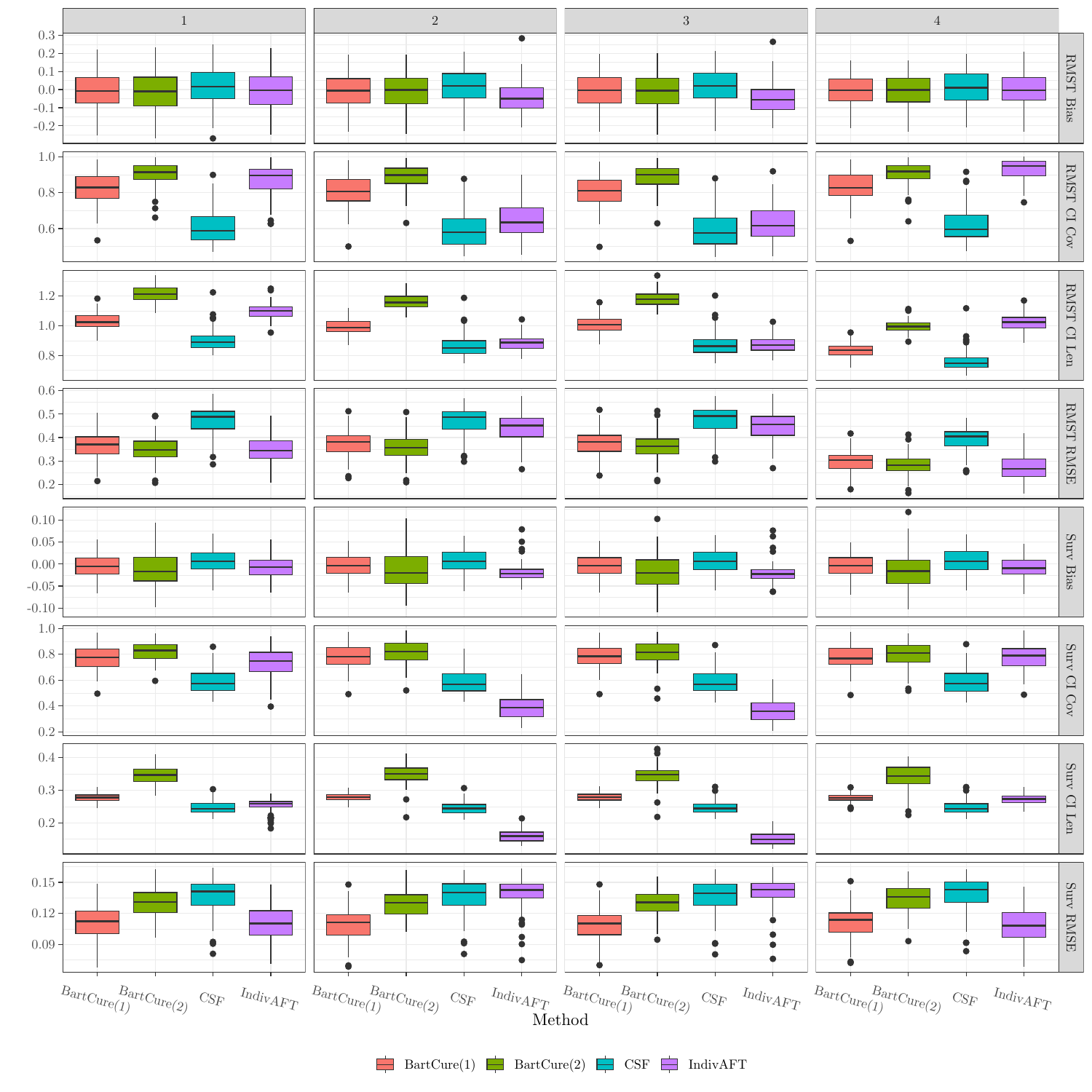}
  \caption{Conditional average treatment effect metrics for the Henderson DGPs \emph{with} a cured fraction.\label{fig:cure_hen}}
\end{figure}

\begin{figure}[p]
  \centering
  \includegraphics[width=\textwidth]{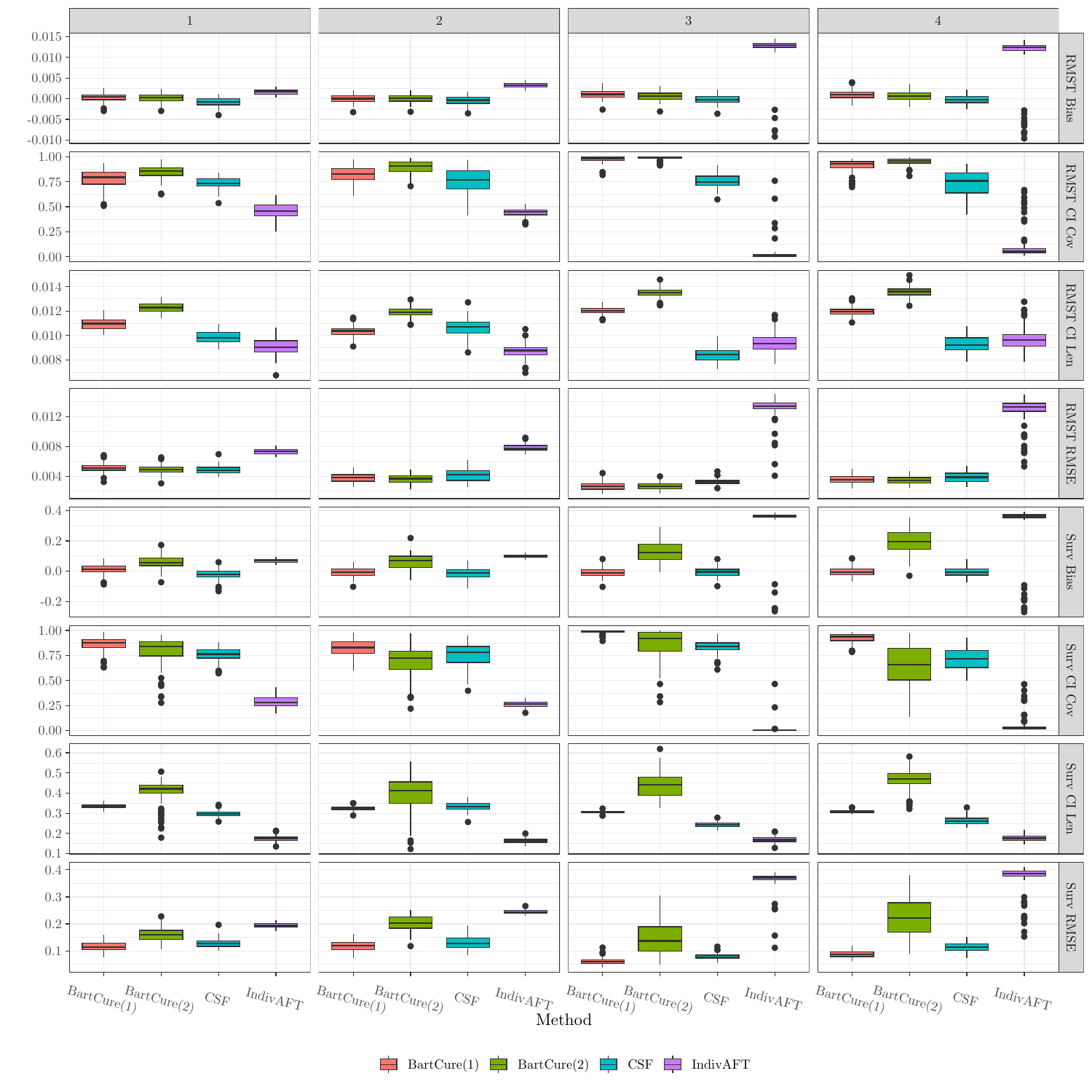}
  \caption{Conditional average treatment effect metrics for the Hu DGPs \emph{with} a cured fraction.\label{fig:cure_hu}}
\end{figure}

\paragraph{Conditional Average Treatment Effects: With a Cured Subpopulation}
Estimating conditional average treatment effects is a much more difficult problem than estimating average treatment effects, and methods differ much more strongly in their performance here. Results for when a cured subpopulation is present are given in Figures~\ref{fig:cure_cui}---\ref{fig:cure_hu}. We did not observe any consistent patterns in terms of overall bias, except that IndivAFT performs very poorly on the Hu settings. For coverage, interval length, and RMSE, \BartCure(1) is the most consistent in the sense that it never performs particularly poorly and always outperforms CSF, but no method attains nominal coverage of confidence/credible intervals; an advantage of \BartCure(1)\ even in this context is that, as shown in Section~\ref{sec:directionality}, it is generally conservative in identifying treatment effect heterogeneity, so in settings where it \emph{does} detect treatment effect heterogeneity it will generally be underestimating it rather than overestimating it.

\begin{figure}[p]
  \centering
  \includegraphics[width=\textwidth]{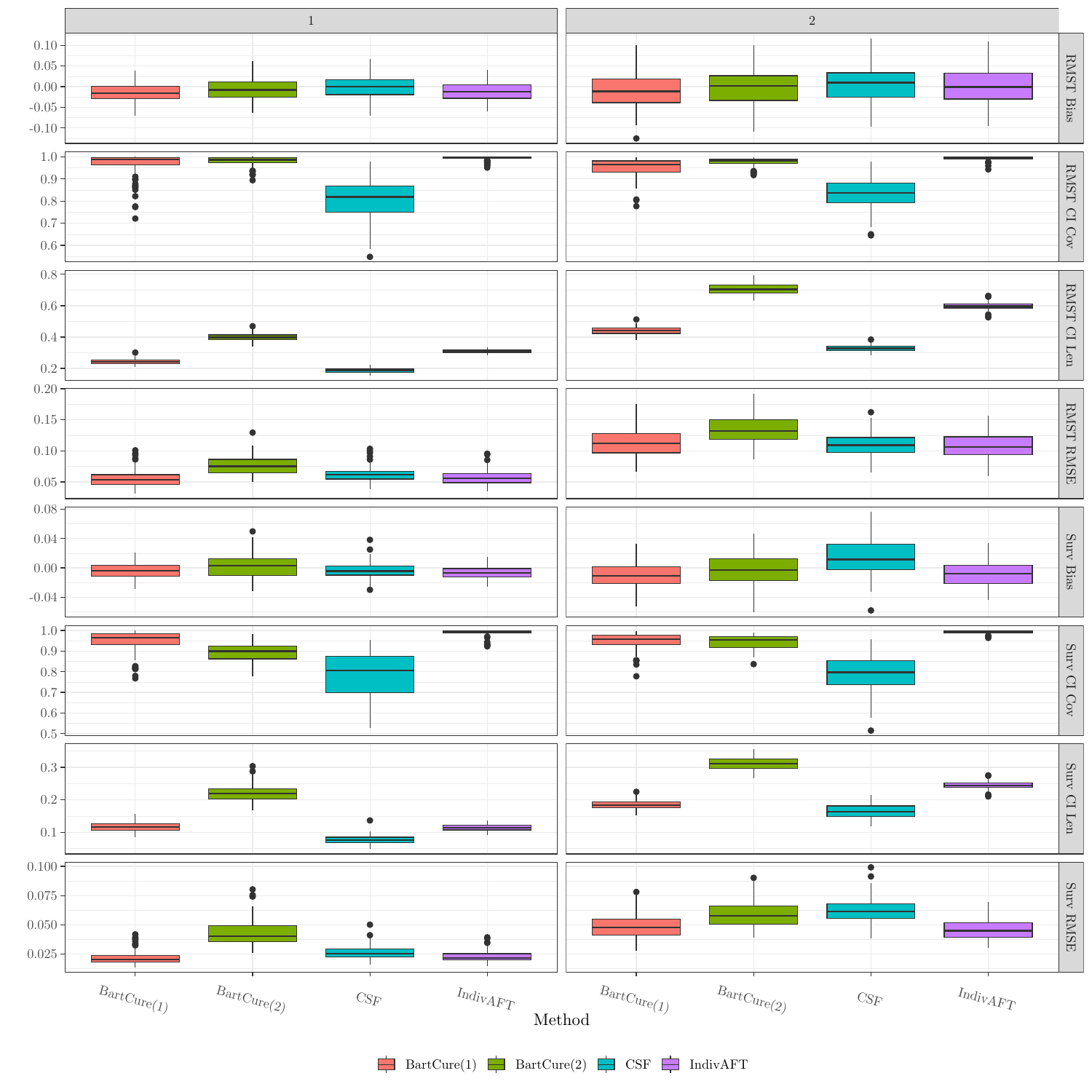}
  \caption{Conditional average treatment effect metrics for the Cui DGPs \emph{without} a cured fraction.\label{fig:sim_cui}}
\end{figure}

\begin{figure}[p]
  \centering
  \includegraphics[width=\textwidth]{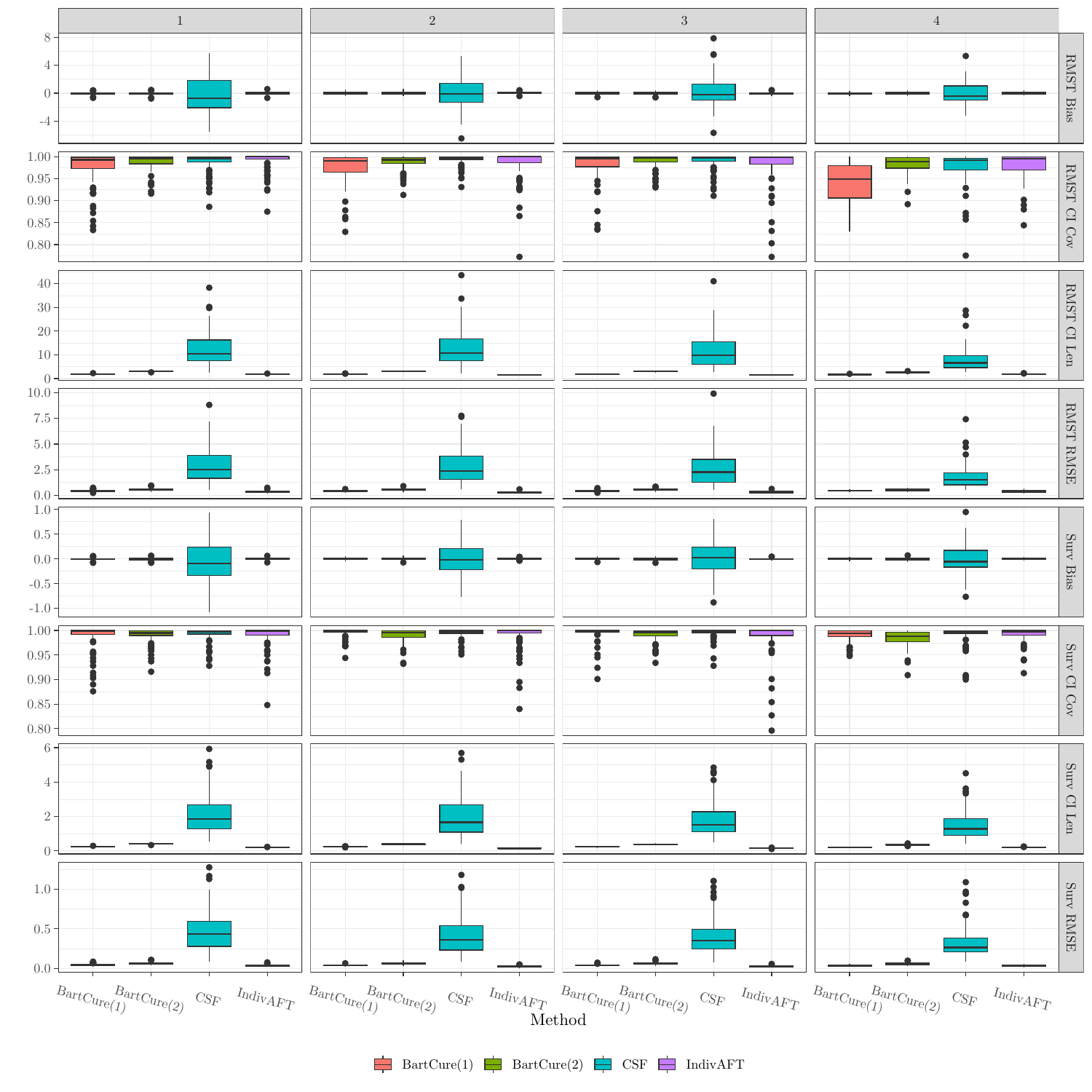}
  \caption{Conditional average treatment effect metrics for the Henderson DGPs \emph{without} a cured fraction.\label{fig:sim_hen}}
\end{figure}

\begin{figure}[p]
  \centering
  \includegraphics[width=\textwidth]{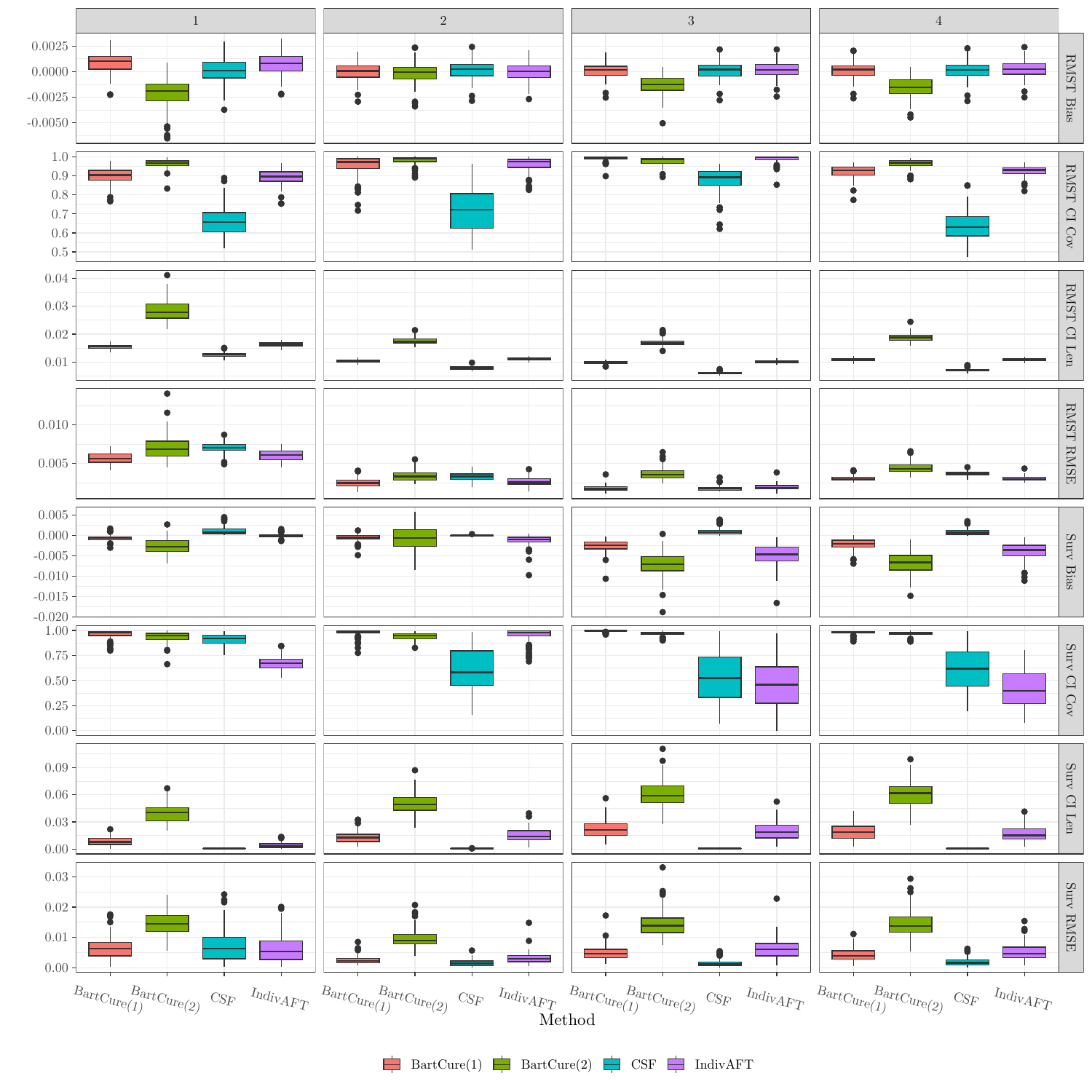}
  \caption{Conditional average treatment effect metrics for the Hu DGPs \emph{without} a cured fraction.\label{fig:sim_hu}}
\end{figure}

\paragraph{Conditional Average Treatment Effects: Without a Cured Subpopulation}
Results are given in Figures~\ref{fig:sim_cui}---\ref{fig:sim_hu}. We again note no consistent patterns in terms of overall bias, and that \BartCure(1) is overall the most robust method and outperforms CSF in terms of RMSE and performance of interval estimands. Interestingly, coverage for all of the effects is much better when there is no cured subpopulation. \BartCure(2)\ remains a more unstable method than \BartCure(1). Just as with the average effects, the CSF breaks down entirely on the Henderson settings, and IndivAFT is mostly competitive with \BartCure(1) except for the Hu~3 and Hu~4 settings. 

\section{Additional Analyses of the CALGB 40101 Trial}
\label{sec:supp-calgb-analyses}

\begin{figure}[t]
  \centering
  \includegraphics[width=0.8\textwidth]{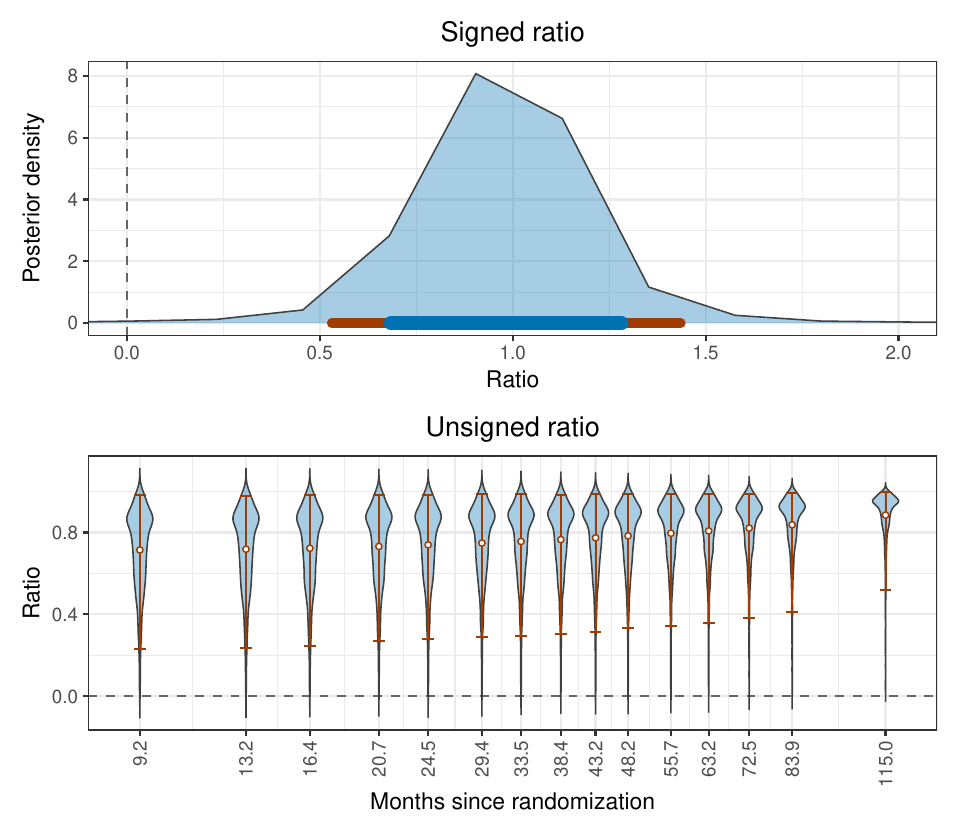}
  \caption{Posterior summaries of the stochastic cure contribution to the RMST effect. The signed ratio is shown at the end of follow-up, where the RMST effect is most stable; the unsigned ratio is shown over time and measures the relative magnitude of the stochastic cure component.}
  \label{fig:pds-stochastic-ratios}
\end{figure}

\paragraph{Proportion of the RMST Effect Attributable to Cure}
Plots of the signed and unsigned shares of the effect attributable to cure as defined in Section~\ref{sec:stochastic-intervention-effects} are given in Figure~\ref{fig:pds-stochastic-ratios}. These plots provide further evidence that the majority of the effect is attributable to the cured subpopulation, and generally there is a substantial amount of uncertainty in the stochastic latency effect.

\begin{figure}[t]
\centering
\includegraphics[width=.8\textwidth]{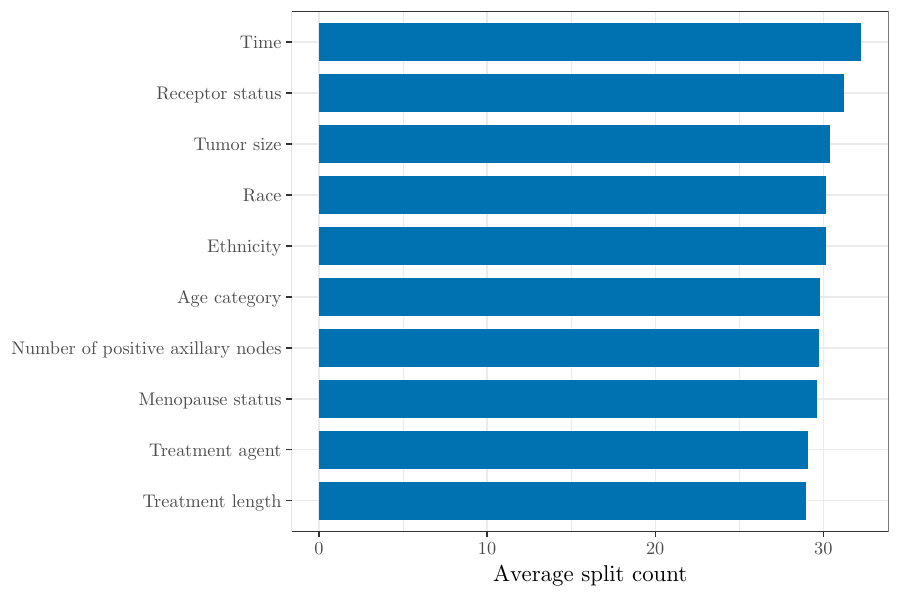}
\caption{Variable importance for the \BartCure\ fit, measured by average split counts across posterior samples.}
\label{fig:pds-variable-importance}
\end{figure}

\paragraph{Variable Importance}
Figure~\ref{fig:pds-variable-importance} displays the variable importances from the \BartCure\ model, quantified using the average number of splitting rules involving each variable as suggested by \citet{chipman2010bart}. The most influential variable in the model was time, which suggests that a proportional hazards variant of the model \eqref{eq:one-forest} that takes $r(t,a,x) = r(a,x)$ would be inadequate, i.e., some variables have a time-varying effect on the hazard. Among baseline clinical covariates, receptor status, tumor size, menopausal status, and age category were the largest contributors to survival. Overall, the variable importance profile suggests that \BartCure\ distributes predictive importance across several clinically relevant covariates rather than concentrating heavily on a single factor. The prominence of receptor status, tumor size, and age is consistent with established prognostic factors in breast cancer survival \citep{abdul2024prognostic}.

\end{document}